\documentclass[aps,superscriptaddress,letterpaper,nofootinbib]{revtex4}

\usepackage{xspace}
\usepackage{graphicx}
\usepackage{epsfig}
\usepackage{float}
\usepackage[nointegrals]{wasysym}
\usepackage{hyperref}
\usepackage{url}
\usepackage{subfig}
\usepackage{amsmath}
\usepackage{xcolor}
\usepackage{amssymb,array}

\newcommand{\be}{\begin{equation}}
\newcommand{\ee}{\end{equation}}
\newcommand{\ba}{\begin{eqnarray}}
\newcommand{\ea}{\end{eqnarray}}

\newcommand{\PH}{H}

\newcommand{\Hboson}{$H$~boson\xspace}
\newcommand{\Hff}{H\!f\!\bar{f}}

\begin{document}

\vspace{0.6cm}

\title{
Snowmass White Paper: 
Prospects of CP-violation measurements \\ with the Higgs boson at future experiments}
\author{Editor: Andrei V. Gritsan \thanks{e-mail: gritsan@jhu.edu}}
\affiliation{Department of Physics and  Astronomy, Johns Hopkins University, Baltimore, MD 21218, USA}
\author{Contributors: Henning Bahl  \thanks{e-mail: hbahl@uchicago.edu}}
\affiliation{University of Chicago, Department of Physics, 5720 South Ellis Avenue, Chicago, IL 60637 USA}
\author{Rahool Kumar Barman \thanks{e-mail: rahool.barmab@okstate.edu}}
\affiliation{Department of Physics, Oklahoma State University, Stillwater, OK, 74078, USA}
\author{Ivanka Bo\v{z}ovi\'{c}-Jelisav\u{c}i\'{c}}{\thanks{ibozovic@vin.bg.ac.rs}} 
\affiliation{``{VIN\u{C}A}" Institute of Nuclear Sciences, University of Belgrade, Belgrade, Serbia} 
\author{Jeffrey Davis  \thanks{e-mail: jdavi231@jhu.edu}}
\affiliation{Department of Physics and Astronomy, Johns Hopkins University, Baltimore, MD 21218, USA}
\author{Wouter Dekens  \thanks{e-mail: wdekens@uw.edu}}
\affiliation{Institute for Nuclear Theory, University of Washington, Seattle WA 91195-1550, USA}
\author{Yanyan Gao \thanks{email: yanyan.gao@ed.ac.uk}}
\affiliation{School of Physics and Astronomy, University of Edinburgh, Edinburgh, EH9 1ES,  UK}
\author{Dorival Gon\c{c}alves \thanks{email: dorival@okstate.edu}}
\affiliation{Department of Physics, Oklahoma State University, Stillwater, OK, 74078, USA}
\author{Lucas~S.~Mandacar\'{u}~Guerra  \thanks{e-mail: lmandac1@jhu.edu}}
\affiliation{Department of Physics and Astronomy, Johns Hopkins University, Baltimore, MD 21218, USA}
\author{Daniel Jeans}{\thanks{daniel.jeans@kek.jp}} 
\affiliation{Institute of Particle and Nuclear Studies, KEK, 305-0801 Tsukuba, Japan} 
\author{Kyoungchul Kong \thanks{email: kckong@ku.edu}}
\affiliation{Department of Physics and Astronomy, University of Kansas, Lawrence, Kansas 66045, USA}
\author{Savvas Kyriacou \thanks{e-mail: skyriac2@jhu.edu}}
\affiliation{Department of Physics and Astronomy, Johns Hopkins University, Baltimore, MD 21218, USA}
\author{Kirtimaan Mohan \thanks{e-mail: kamohan@msu.edu}}
\affiliation{Department of Physics and Astronomy, Michigan State University, East Lansing, MI 48824, USA}
\author{Ren-Qi Pan  \thanks{e-mail: renqi.pan@cern.ch}}
\affiliation{Zhejiang Institute of Modern Physics, Department of Physics, Zhejiang University, Hangzhou, 310027, P. R. China}
\author{Jeffrey Roskes \thanks{e-mail: hroskes@jhu.edu}}
\affiliation{Department of Physics and Astronomy, Johns Hopkins University, Baltimore, MD 21218, USA}
\author{Nhan V. Tran  \thanks{e-mail: ntran@fnal.gov}}
\affiliation{Fermi National Accelerator Laboratory (FNAL), Batavia, IL 60510, USA}
\author{Natasa Vuka\u{s}inovi\'{c} \thanks{nvukasinovic@vin.bg.ac.rs}}
\affiliation{``{VIN\u{C}A}" Institute of Nuclear Sciences, University of Belgrade, Belgrade, Serbia} 
\author{Meng Xiao  \thanks{e-mail: meng.xiao@cern.ch}}
\affiliation{Zhejiang Institute of Modern Physics, Department of Physics, Zhejiang University, Hangzhou, 310027, P. R. China}
%

%
%

\date{May 16, 2022}

\begin{abstract}
\vspace{2mm}
The search for $CP$ violation in interactions of the Higgs boson with either fermions or bosons provides 
attractive reference measurements in the Particle Physics Community Planning Exercise (a.k.a. ``Snowmass"). 
Benchmark measurements of $CP$ violation provide a limited and well-defined set of parameters
that could be tested at  the proton, electron-positron, photon, and muon colliders, and compared to 
those achieved through study of virtual effects in electric dipole moment measurements. 
We review the current status of these $CP$-sensitive studies and provide projections to future measurements.
\end{abstract}
\maketitle

\thispagestyle{empty}

\clearpage


\tableofcontents
\clearpage


\section{Introduction}
\label{sect:cp_intro}

The search for $CP$ violation is an important research direction of future experiments in particle physics. 
$CP$ violation is one of the requirements for baryogengesis~\cite{Sakharov:1967dj}. So far the only experimental 
evidence for $CP$ violation comes from quark flavor physics, which is consistent with the CKM mechanism appearing
in the Standard Model (SM) of particle physics~\cite{Kobayashi:1973fv}. 
This SM mechanism is believed to be insufficient for generating the observed predominance 
of baryon matter over antimatter on a cosmological scale~\cite{Shaposhnikov:1987tw}. 
Therefore, the search for $CP$ violation in interactions of the Higgs boson ($H$) with 
either fermions or bosons is an interesting path to search for a new mechanism. 

Through the study of the $HVV$ and $\Hff$ tensor structure of interactions of the \Hboson with vector bosons
($Z, W, \gamma,$ g) and fermions ($t$, $\tau$), the CMS and ATLAS experiments on LHC have established that 
the $J^{PC}$ quantum numbers of the \Hboson should be $0^{++}$, if this boson has definite $P$ and $C$~\cite{Chatrchyan:2012jja,
Aad:2013xqa,Chatrchyan:2013mxa,Khachatryan:2014kca,Aad:2015mxa,Khachatryan:2016tnr,
Aad:2016nal,Sirunyan:2017tqd,Aaboud:2017oem,Aaboud:2017vzb,Aaboud:2018xdt,Sirunyan:2019twz,
Sirunyan:2019nbs,Sirunyan:2020sum,ATLAS:2020ior,Aad:2020mnm,Sirunyan:2021fpv,CMS:2021sdq,
ATLAS:2021pkb,CMS:2022ley,Collaboration:2022mlq,CMS-PAS-HIG-21-006,ATLAS-CONF-2022-016,ATLAS-CONF-2022-032}. 
This observation is consistent with the  Standard Model (SM) expectation for these quantum numbers to be that of 
the vacuum. However, small violation of $CP$ symmetry in those interactions cannot be excluded within the experimental 
precision of current measurements. Squeezing the allowed range of $CP$-violating parameters, or, alternatively, 
discovering non-zero $CP$ violation in the \Hboson interactions, becomes an important target of experimental
measurements~\cite{Nelson:1986ki,Soni:1993jc,Chang:1993jy,Barger:1993wt,
Arens:1994wd,BarShalom:1995jb,Gunion:1996xu,Han:2000mi,Plehn:2001nj,Choi:2002jk,Buszello:2002uu,Hankele:2006ma,
Accomando:2006ga,Godbole:2007cn,Hagiwara:2009wt,Gao:2010qx,DeRujula:2010ys,Christensen:2010pf,
Gainer:2011xz,Bolognesi:2012mm,Ellis:2012xd,Chen:2012jy,Gainer:2013rxa,Artoisenet:2013puc,Anderson:2013afp,Chen:2013waa,
Chen:2013ejz,Gainer:2014hha,Gonzalez-Alonso:2014eva,Dolan:2014upa,Demartin:2014fia,Chen:2014gka,Buckley:2015vsa,
Greljo:2015sla,Gritsan:2016hjl,deFlorian:2016spz,Hartmann:2015aia,Dawson:2018pyl,Dedes:2018seb,Dawson:2018liq,
Brivio:2019myy,Gritsan:2020pib,Martini:2021uey,Davis:2021tiv,Barman:2021yfh,Barman:2022pip}.

Future high-energy physics experiments, either planned or proposed, have unique features for testing $CP$ violation
in the \Hboson interactions. For example, photon and muon colliders with a beam polarization scan
could provide a unique opportunity to search for $CP$ violation in couplings to either photons or muons. 
The electron-positron collider is positioned uniquely to search for $CP$ violation in $HVV$ interactions, with vector bosons 
appearing in electron-positron annihilation, and allow for $CP$ studies in decay. Proton colliders provide an array 
of opportunities for $HVV$ and $\Hff$ studies in both production and decay, as already demonstrated by the LHC experiments. 

This makes $CP$ violation studies with the \Hboson an attractive reference measurement in the Particle Physics 
Community Planning Exercise (a.k.a. ``Snowmass"), organized by the US High Energy Physics community
to set directions in the field of particles physics for the next decade and beyond. Benchmark measurements of 
$CP$ violation in interactions of the \Hboson with SM particles provide a limited and well-defined set of parameters
that could be tested at future high-energy physics colliders. Moreover, these measurements can also be achieved 
through study of virtual effects in quark flavor physics and electric dipole moment (EDM) measurements. 
These $CP$ violation effects are tiny in the SM, and they therefore become excellent null tests for comparing 
performance of future facilities. Beyond-the-SM (BSM) theories predict sizable $CP$ violation effects, which could 
have profound implications for the future of particle physics, if discovered.

\section{Framework of the Higgs $CP$ study for Snowmass}
\label{sect:cp_framework}

Since the discovery of the \Hboson by the ATLAS and CMS experiments on the Large Hadron Collider 
(LHC)~\cite{Aad:2012tfa,Chatrchyan:2012xdj}, the search for $CP$ violation in its interactions started 
immediately~\cite{Chatrchyan:2012jja,Aad:2013xqa,Chatrchyan:2013mxa}. The $CP$ violation parameters 
were identified as benchmark measurements in the  Snowmass-2013 Particle Physics Community Planning 
Exercise~\cite{Dawson:2013bba}. In that study, $CP$-violating parameters were defined in the coupling
of the \Hboson to massive vector bosons ($HVV$), to massless vector bosons 
($H\gamma\gamma$, $HZ\gamma$ , $H$gg), and to fermions ($Ht\bar{t}$, $H\tau\tau$, $H\mu\mu$). 
In this work, we build on that study, take advantage of both experimental and theoretical progress
in the study of the \Hboson interactions over the past decade, and make assessment of prospects
of the \Hboson $CP$ study at the future facilities, both proposed and planned. 

On the experimental front, measurements of $CP$ violation parameters have been achieved on LHC
experiments~\cite{Chatrchyan:2012jja,
Aad:2013xqa,Chatrchyan:2013mxa,Khachatryan:2014kca,Aad:2015mxa,Khachatryan:2016tnr,
Aad:2016nal,Sirunyan:2017tqd,Aaboud:2017oem,Aaboud:2017vzb,Aaboud:2018xdt,Sirunyan:2019twz,
Sirunyan:2019nbs,Sirunyan:2020sum,ATLAS:2020ior,Aad:2020mnm,Sirunyan:2021fpv,CMS:2021sdq,
ATLAS:2021pkb,CMS:2022ley,Collaboration:2022mlq}. 
This allows us to make realistic quantitative projections to the HL-LHC.
The main change on the theoretical front has been development of the Effective Field Theory (EFT)
framework, and in particular SMEFT, where $CP$ violation naturally appears from a sub-set of 
higher-dimension operators. Both developments have been discussed within the LHC Higgs and 
LHC EFT Working Groups~\cite{lhcxwg,lhceft}, where we rely on some of their efforts. 

One could consider the study of $CP$ violation to be redundant with respect to the larger
project of EFT global fits. However, there are two reasons which make this consideration less
reliable for the Snowmass studies. First of all, the global EFT fits are very complex with many parameters
and assumptions invoked to reduce the number of those parameters. One of the common constraints
applied in the current global EFT fits is the lack of $CP$-odd operators, effectively setting all 
$CP$ violation effects to zero. Second, even when such $CP$ constraints are not invoked, the 
actual measurements are often based on experimental information which is not necessarily 
truly $CP$-sensitive. This means that the measurement may be sensitive to the presence of 
the higher-dimension operators, but may not distinguish well between $CP$-odd and $CP$-even terms. 
Therefore, the idea of the dedicated $CP$-sensitive measurements of the \Hboson for the Snowmass 
studies is to provide simple but at the same time reliable benchmarks which could serve as a guide 
to compare future facilities. 

As an example of the future measurement projections based on the global EFT fits, let us refer to the
\Hboson studies performed for the 2020 European Strategy for Particle Physics Update~\cite{deBlas:2019rxi}.
In the EFT description of the \Hboson couplings, either 18 (with flavor universality) or
30 (with neutral diagonality) $CP$-even operators in the so-called Higgs basis were considered 
within the SMEFT framework, which invokes the SU(2)$\times$U(1) EW symmetry. 
To assess the sensitivity to deviations from the SM in a basis-independent way 
the results of the fit were projected onto the following \Hboson effective couplings:
\begin{equation}
{\rm g}^{\rm eff~2}_{HX}\equiv
\frac{\Gamma_{H\to X}}{\Gamma^{\rm SM}_{H\to X}}
\,.
\label{eq:gHX} 
\end{equation}
These parameters are convenient to compare different studies in a straightforward manner. 
However, these parameters do not allow for the $CP$ structure in the $HX$ interaction. 
Therefore, we expand this set of $CP$-conserving parameters with the following set, 
allowing for $CP$ violation in each $HX$ interaction:
\begin{equation}
f_{CP}^{HX}\equiv
\frac{\Gamma^{CP\rm\,odd}_{H\to X}
}{\Gamma^{CP\rm\,odd}_{H\to X}
+\Gamma^{CP\rm\,even}_{H\to X}}
\,,
\label{eq:fCP} 
\end{equation}
where the partial decay $H\to X$ width is calculated with either the $CP$-odd or $CP$-even part of the amplitude. 
This definition is consistent with the $CP$-sensitive parameters $f_{CP}$ defined for the Snowmass-2013
study~\cite{Dawson:2013bba}. These $f_{CP}$ parameters have been adopted in the LHC measurements
as well, for a recent summary refer to Ref.~\cite{Sirunyan:2021fpv}. 
Therefore, we adopt Eq.~(\ref{eq:fCP}) for the benchmark parameter measurements. 

We note that Eq.~(\ref{eq:fCP}) is defined in decay of the \Hboson. For example, 
the general scattering amplitude that describes the interaction of the \Hboson 
with the fermions, such  as $\tau^+\tau^-$, $\mu^+\mu^-$,  $b\bar{b}$, and $t\bar{t}$,
can be written as
%
\begin{eqnarray}
&& A( H \to f\bar f) = \frac{m_f}{v}
\bar u_2 \left ( b_{1}^{\Hff} + i b_{2}^{\Hff} \gamma_5 \right ) u_1
\,.
\label{eq:ampl-spin0-qq}
\end{eqnarray}
%
Therefore, the $CP$-sensitive parameter takes the form
\begin{equation}
f_{CP}^{\Hff}\equiv
\frac{|b_2^{\Hff}|^2}{|b_1^{\Hff}|^2+|b_2^{\Hff}|^2} = \sin^2\left( \alpha^{\Hff} \right)
\,.
\label{eq:fHff} 
\end{equation}
Technically, Eq.~(\ref{eq:fCP}) does not cover $Ht\bar{t}$ interactions, because the decay $H\to t\bar{t}$ is not possible. 
However, we expand the definition in Eq.~(\ref{eq:fHff}) to all fermion couplings. The effective mixing angle $\alpha^{\Hff}$,
introduced in Eq.~(\ref{eq:fHff}), is often used in describing the $CP$-odd amplitude contribution. However, we adopt a
more general parameterization with effective cross-section fractions because they allow more than two amplitude 
contributions, as this becomes important in description of the $HVV$ interactions, discussed below. 

For the coupling to the gauge bosons, such as $WW$, $ZZ$, $Z\gamma$, $\gamma \gamma$, or gg,
the scattering amplitude can be written as 
\begin{equation}
A(H \to V_1V_2) = v^{-1} \left ( 
  a^{HVV}_{ 1} m_{V}^2 \epsilon_1^* \epsilon_2^* 
+ a^{HVV}_{ 2} f_{\mu \nu}^{*(1)}f^{*(2),\mu \nu}
+ a^{HVV}_{ 3}  f^{*(1)}_{\mu \nu} {\tilde f}^{*(2),\mu  \nu}
\right )\,,
\label{eq:fullampl-spin0} 
\end{equation}
where $a_i^{HVV}$ are generally $q^2$-dependent coefficients scaling the three unique Lorentz structures,
described with the help of the (conjugate) field strength tensor $f^{(i),\mu \nu}$ (${\tilde f}^{(i),\mu  \nu}$)
of a gauge boson with momentum $q_i$ and polarization vector $\epsilon_i$. 
In the following, we will keep only the first-order $q^2$-expansion of Eq.~(\ref{eq:fullampl-spin0})
with constant coefficients $a_{i}$,
which correspond to dimension-six operators in the effective Lagrangian formulation. 
The presence of the $CP$-odd contribution $a^{HVV}_{3}$, which can be treated as constant
in this expansion, indicates $CP$ violation, and the $CP$-sensitive parameter takes the form
%
\begin{eqnarray}
&& f_{CP}^{HVV} =  \frac{|a_{3}^{HVV} |^2}{\sum |a_{i}^{HVV} |^2 (\sigma^{HVV} _i/\sigma^{HVV} _3)} \,,
\label{eq:fractions}
\end{eqnarray}
%
where $\sigma_i$ is the effective cross-section of the $H\to VV$
decay process corresponding to $a_{i}=1, a_{j \ne i}=0$.

\begin{table}[ht]
\renewcommand{\arraystretch}{1.5}
\captionsetup{justification=centerlast}
\caption{
List of expected precision (at 68\% C.L.) of ${CP}$-sensitive measurements of the parameters $f_{CP}^{HX}$ defined in Eq.~(\ref{eq:fCP}).
Numerical values are given where reliable estimates are provided, 
$\checked$ mark indicates that feasibility of such a measurement could be considered.
The $e^+e^-\to ZH$ projections are performed with $Z\to\ell\ell$ in Appendix~\ref{app:B} 
but scaled to a ten times larger luminosity to account for $Z\to q\bar{q}$. 
}
\vspace{-0.4cm}
\begin{center}
\begin{tabular}{|l|ccccccccccc|c|}
\hline\hline
Collider                     &   $pp$       &   $pp$    &   $pp$    &   $e^+e^-$    &   $e^+e^-$    &  $e^+e^-$    &  $e^+e^-$    & $e^-p$ & $\gamma\gamma$ &  $\mu^+\mu^-$  & $\mu^+\mu^-$  & target \\
E (GeV)                     &   14,000   &   14,000  &   100,000     &  250          &  350                 & 500             &   1,000       &   1,300  &       125               &    125          & 3,000                        &   (theory) \\
${\cal L}$ (fb$^{-1}$) & 300  & 3,000  & 30,000 & 250          &  350                  & 500              &    1,000       &    1,000        &        250        &      20      &  1,000 &     \\ 
\hline
\hline
$HZZ/HWW$ &  $4.0\!\cdot\!10^{-5}$ &  $2.5\!\cdot\!10^{-6}$ & \checked & $3.9\!\cdot\!10^{-5}$ & $2.9\! \cdot\!10^{-5}$ & $1.3\!\cdot \!10^{-5}$ & $3.0\!\cdot\!10^{-6}$  & \checked  & \checked  &  \checked  &  \checked  & $<10^{-5}$  \\
\hline
\hline
  $H\gamma\gamma$ & --  & 0.50  & \checked & -- & -- & -- & -- & -- & 0.06 & -- & --  & $<10^{-2}$  \\
\hline
     $HZ\gamma$ & --  & $\sim\!1$  & \checked &-- & -- & -- & $\sim\!1$   & -- & -- & --  & --  & $<10^{-2}$   \\
\hline
  $H$gg & $0.12$ & $0.011$  & \checked & -- & -- & -- & -- & -- & -- & -- &  -- & $<10^{-2}$   \\
\hline
\hline
 $Ht\bar{t}$ & 0.24  & 0.05  & \checked & -- & -- & 0.29 & 0.08  & \checked  & -- & --  & \checked & $<10^{-2}$  \\
\hline
 $H\tau\tau$ &  0.07 & 0.008  & \checked & $0.01$ & $0.01$  &  ${0.02}$  & ${0.06}$  &  -- & \checked & \checked & \checked & $<10^{-2}$  \\
\hline
 $H\mu\mu$ & --  & -- & -- & -- & -- & -- & --  & -- & -- & \checked & -- & $<10^{-2}$   \\
\hline
\hline
\end{tabular}
\end{center}
\label{table-cpscenarios}
\end{table}

This brings us to the summary of possible $CP$-sensitive measurements in the \Hboson interactions in Table~\ref{table-cpscenarios}. 
In the following, we will review unique features of photon, muon, hadron, and electron-positron colliders, where we
keep the energy and luminosity scenarios the same as in the Snowmass-2013 studies~\cite{Dawson:2013bba} for easy 
comparison and because several projections (such as $CP$ violation in fermion couplings) have not been updated. 
More recent projections for electron-positron colliders have been shown with higher luminosity,
and we indicate in Appendix~\ref{app:B} how expectations scale to an order of magnitude higher luminosity
for some of the couplings. These higher luminosity collider scenarios are consistent within a factor of two with 
the more recent recommendations for Snowmass-2022 studies, as outlined in Ref.~\cite{Dawson:2022zbb}. 


\section{Prospects of Higgs $CP$ measurements at a photon collider}
\label{sect:photon}

The photon collider has a unique feature in that it can be used to study the \Hboson couplings to photons in direct production 
$\gamma\gamma\to H$. It is also possible to study the \Hboson couplings in decay, such as $CP$ structure
in $H\to \tau^+\tau^-$ or $H\to 4f$. However, the decay measurements critically depend on the number 
of produced $H$ bosons, and a Higgs factory in either lepton or proton collisions is better positioned 
to make those measurements. In this Section, therefore, we focus on the $H\gamma\gamma$ measurements, 
which are unique to the photon collider. 

The coupling of the \Hboson to two photons cannot happen at tree level, but can be generated 
by loops of any charged particles. In the SM, those are the charged fermions and W boson. 
In the SM, $CP$ violation is tiny, as it can be generated only at three-loop level.
In BSM theories, new heavy states can contribute to the loop, and could generate sizable $CP$ violation. 
Alternatively, $CP$ violation in the \Hboson couplings to SM particles could also generate $CP$-odd
contributions to the $H\gamma\gamma$ loop. 
Both $H\to\gamma\gamma$ decay and $\gamma\gamma\to H$ production can be
parameterized with the $CP$-even $a_2^{H\gamma\gamma}$ and $CP$-odd $a_3^{H\gamma\gamma}$
contributions in Eq.~(\ref{eq:fullampl-spin0})
with the ratio $\sigma^{H\gamma\gamma} _2/\sigma^{H\gamma\gamma} _3=1$ in Eq.~(\ref{eq:fractions}). 
However, without access to the photon polarization, it is not possible to distinguish between the two
contributions in the $H\to\gamma\gamma$ decay.\footnote{An attempt to measure photon polarization 
in its conversion is possible~\cite{Bishara:2013vya}, but it suffers from a significant loss of statistical precision. We will discuss 
the photon polarization measurements in the $H\to\gamma^*\gamma^*\to 4f$ process in Section~\ref{sect:hadron}.}
Therefore, variation of the photon polarization in the photon collider becomes a unique approach to 
study the $CP$ structure of the $H\gamma\gamma$ vertex. 

Three parameters ${\cal A}_1, {\cal A}_2, {\cal A}_3$ sensitive to $C\!P$ violation
have been defined in the context of the photon collider~\cite{Grzadkowski:1992sa, Kramer:1993jn, Gunion:1994wy}.
The ${\cal A}_1$ parameter can be measured as an asymmetry in the \Hboson production cross-section
between the $A_{++}$ and $A_{--}$ circular polarizations of the beams. This asymmetry 
is the easiest to measure, but it is proportional to $\Im{m}(a_2^{H\gamma\gamma}a_3^{H\gamma\gamma~*})$ and is zero
when $a_2^{H\gamma\gamma}$ and $a_3^{H\gamma\gamma}$ are real,
as expected for the two loop-induced couplings with heavier particles in the loops. 
A more interesting parameter,
%
\begin{eqnarray}
{\cal A}_3 
= \frac{|A_{\parallel}|^2-|A_{\perp}|^2}{|A_{\parallel}|^2+|A_{\perp}|^2}
= \frac{2\Re{e}(A_{--}^*A_{++})}{|A_{++}|^2+|A_{--}|^2}
= \frac{|a_2^{H\gamma\gamma}|^2-|a_3^{H\gamma\gamma}|^2}{|a_2^{H\gamma\gamma}|^2+|a_3^{H\gamma\gamma}|^2}
= (1-2f_{CP}^{H\gamma\gamma}),
\label{eq:photoncp}
\end{eqnarray}
%
can be measured as an asymmetry between two configurations with the linear
polarization of the photon beams, one with parallel and the other with orthogonal 
polarizations. 

In Ref.~\cite{Asner:2001ia}, a careful simulation of the process has been performed. 
The degree of linear polarization at the maximum energies is 60\% for an electron
beam of energy $E_0 \approx 110$\,GeV and a laser wavelength $\lambda \approx 1\, \mu{m}$.
The expected uncertainty on ${\cal A}_3$ is 0.11 for $2.5\cdot 10^{34} \times
10^7$ = 250\,fb$^{-1}$ integrated luminosity and $m_H=120$\,GeV. 
This translates to a $f_{CP}^{H\gamma\gamma}$ uncertainty of 0.06, 
which we enter as an estimate in Table~\ref{table-cpscenarios}. 

The $CP$ mixture study at a photon collider was also shown based on a sample of 50,000 raw
$\gamma\gamma\to H$ events assuming 80\% circular polarization of both electron beams~\cite{Chou:2013xaa}.  
This study corresponds to a ${\cal A}_1$ asymmetry measurement, 
with expected precision on ${\cal A}_1$ of about 1\%. 
However, this asymmetry is expected to be zero with real coupling constants $a_2^{H\gamma\gamma}$ and 
$a_3^{H\gamma\gamma}$ and is therefore of limited interest compared to $f_{CP}^{H\gamma\gamma}$.


\section{Prospects of Higgs $CP$ measurements at a muon collider}
\label{sect:muon}

Similarly to the photon collider, we focus on a unique feature of the muon collider
operating at the \Hboson pole. This allows one to measure the $CP$ structure of the $H\mu\mu$ vertex 
with the beam polarization in the $\mu^+\mu^-\to H$ process.
It is not possible to study the $CP$ structure in the $H\to \mu^+\mu^-$ decay because the muon polarization is not accessible.
The muon collider may become the only facility allowing a measurement of $CP$ structure
in the \Hboson's connection to the second-family fermions.
At a muon collider operating both at the \Hboson pole and at higher energy, 
analysis of the \Hboson decays is also possible. However, this analysis is similar to 
the studies performed at other facilities and depends critically on the number of the $H$ bosons
produced and their purity. 

\subsection{Muon collider at the \Hboson pole}
\label{sect:mhmuon}

At a muon collider operating at the resonance pole, the $CP$ quantum numbers of the states can be determined 
if the muon beams can be transversely polarized.  
The cross section for production of a resonance takes the form~\cite{Grzadkowski:2000hm}
\begin{equation}
  \sigma_{\rm pol}(\zeta) = \sigma_{\rm unpol} \left( 1 + P_L^+ P_L^-
  + P_T^+ P_T^- \left[ \frac{(b_1^{H\mu\mu})^2 - (b_2^{H\mu\mu})^2}{(b_1^{H\mu\mu})^2 + (b_2^{H\mu\mu})^2} \cos\zeta
    - \frac{2b_1^{H\mu\mu}b_2^{H\mu\mu}}{(b_1^{H\mu\mu})^2 + (b_2^{H\mu\mu})^2} \sin\zeta \right] \right),
\end{equation}
which depends on $P_T$ ($P_L$), the degree of transverse (longitudinal)
polarization of each of the beams and $\zeta$ is the angle of the
$\mu^+$ transverse polarization relative to that of the $\mu^-$
measured using the direction of the $\mu^-$ momentum as the $z$ axis.
In particular, muon beams polarized in the same transverse direction
selects out the $C\!P$-even state, while muon beams polarized in opposite
transverse directions (i.e., with spins $+1/2$ and $-1/2$ along one
transverse direction) selects out the $C\!P$-odd state. 
A quantitative estimate of the muon collider precision in the measurement of 
$CP$ structure of the $H\mu\mu$ vertex is left for future studies, which we indicate
with a checkmark in Table~\ref{table-cpscenarios}. 

\subsection{High-energy muon collider}
\label{sect:hemuon}

Operation of the muon collider at higher energies will allow access to associated production
of the \Hboson and study $CP$ properties in those processes. Such studies would 
be similar to those discussed in Sections~\ref{sect:hadron} and~\ref{sect:lepton}
and would depend on achieved performance of the muon collider.
At energies around 1\,TeV, VBF production of the \Hboson dominates, similarly to the $e^+e^-$ collider. 
The dominant channel $\mu^+\mu^-\to \nu_\mu\bar\nu_\mu(W^+W^-)\to \nu_\mu\bar\nu_\mu H$
does not provide kinematic information to analyze the final state with missing neutrinos,
but provides $H$ bosons for analysis of their decay. 
The momentum of the \Hboson in this VBF production provides sensitivity to the higher-dimension 
operators, but does not allow one to separate $CP$-odd and $CP$-even contributions. 
The other channel $\mu^+\mu^-\to \mu^+\mu^-(ZZ/Z\gamma^*/\gamma^*\gamma^*)\to \mu^+\mu^- H$
provides sufficient information to analyze potential $CP$ structure in the $HZZ/HZ\gamma/H\gamma\gamma$ couplings. 
The $t\bar{t}H$ production allows access to $CP$ in the $Ht\bar{t}$ coupling, which is accessible at energies above 0.5\,TeV. 
It is pointed out in Ref.~\cite{Barman:2022pip} that at energies around 10\,TeV, VBF production of $t\bar{t}H$ and $t\bar{q}H$ 
becomes important. 
According to Ref.~\cite{Barman:2022pip}, it is expected to achieve constraints on $f_{CP}^{Ht\bar{t}}<0.67, 0.024$, and~$0.003$ in three scenarios
of the muon collider at 1\,TeV with 0.1\,ab$^{-1}$ of data, 10\,TeV with 10\,ab$^{-1}$, and 30\,TeV with 10\,ab$^{-1}$. 
We do not enter numerical values in Table~\ref{table-cpscenarios} due to uncertain muon collider scenarios,
but point to the possible measurements with the checkmarks. 


\clearpage

\section{Prospects of Higgs $CP$ measurements at a hadron collider}
\label{sect:hadron}

Hadron colliders provide essentially the full spectrum of possible measurements sensitive to $CP$ violation
in the \Hboson interactions, as outlined in Table~\ref{table-cpscenarios}, with the exception of the $H\mu\mu$ vertex.
Here we discuss applications to the LHC experiment and its high-luminosity upgrade (HL-LHC), with a proton-proton 
collision energy around 14\,TeV. There is a proposal for a 100\,TeV proton-proton collider (FCC-hh or SPPC) 
which is designed to collect a total luminosity of 20\,ab$^{-1}$. Given increase in the $H$ boson production cross 
sections by more than an order of magnitude in most production channels, we would expect about two orders of magnitude 
more $H$ bosons produced and the corresponding increase in precision on $f_{CP}^{HX}$ by between one and two orders 
of magnitude, given that these measurements are statistics limited, when compared to HL-LHC. 
Therefore, we expect the 100\,TeV $pp$ collider to surpass all other experiments, but we leave only checkmarks
in such cases in Table~\ref{table-cpscenarios} because detailed simulation studies have not been performed yet
and the proposed timescale of this experiment is longer than in most other cases. 


\subsection{Gluon fusion process at a hadron collider}
\label{sect:ggH}

\begin{figure}[b]
\centerline{
\setlength{\epsfxsize}{0.31\linewidth}\leavevmode\epsfbox{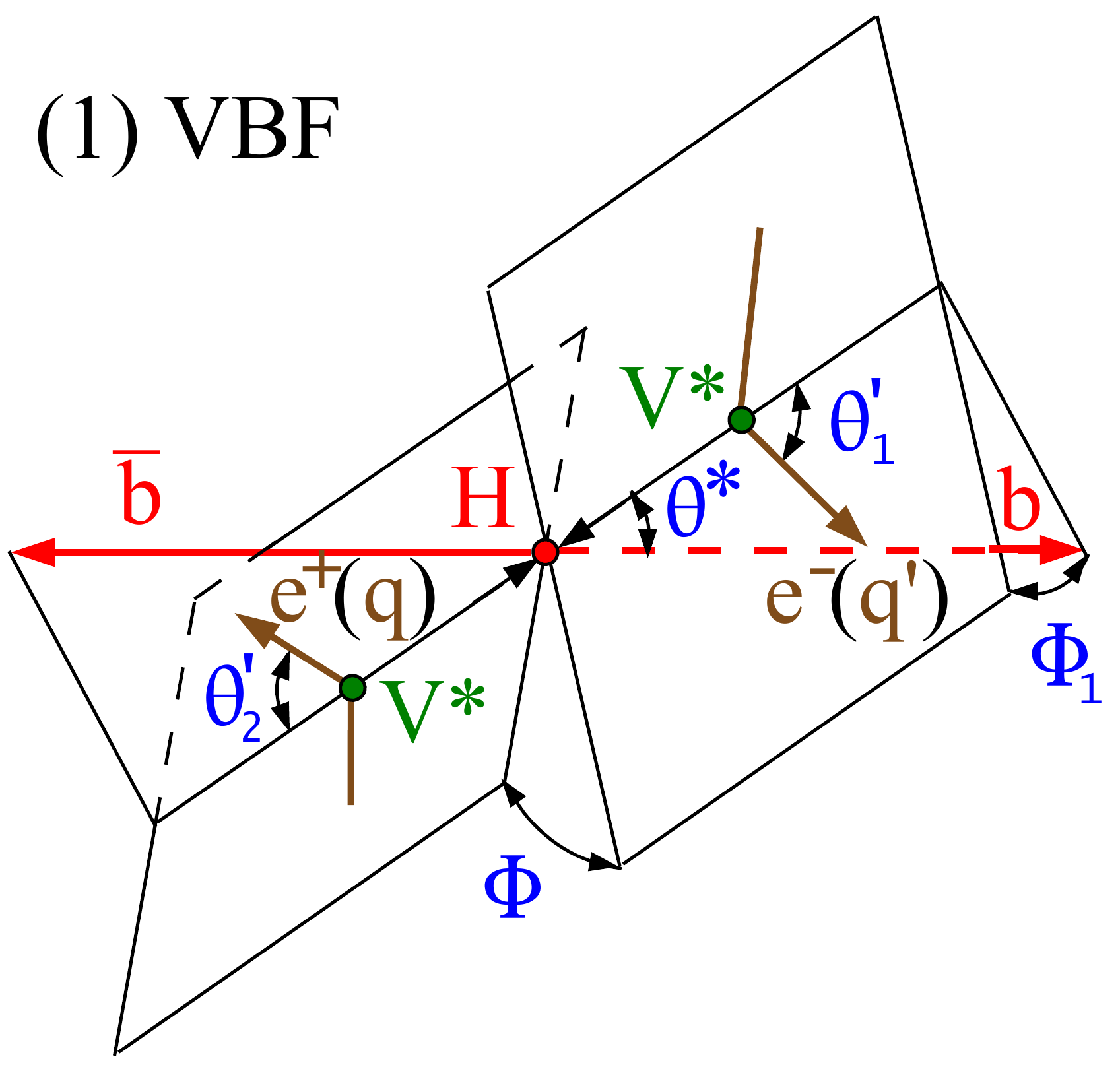}
\setlength{\epsfxsize}{0.31\linewidth}\leavevmode\epsfbox{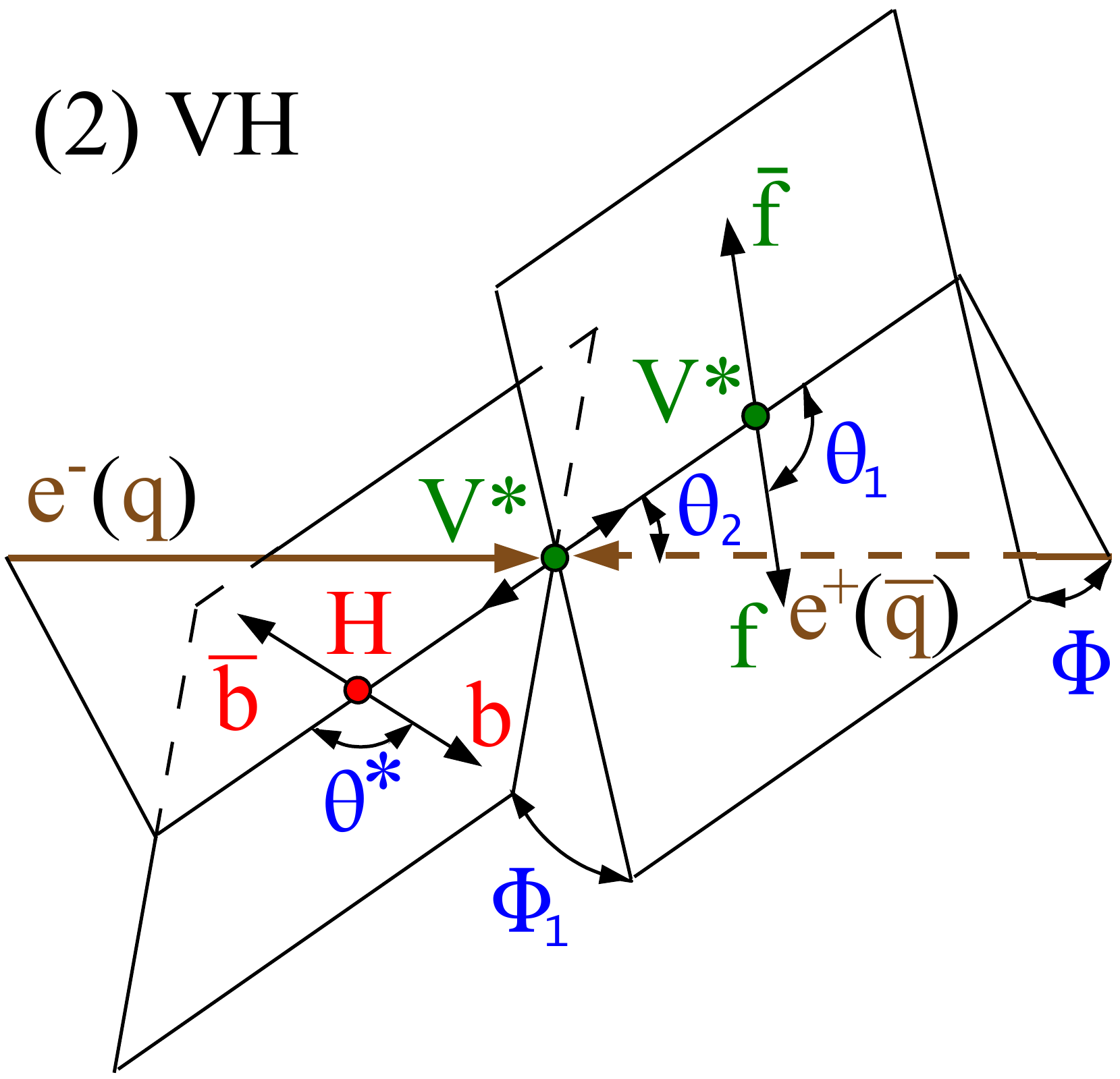}
\setlength{\epsfxsize}{0.31\linewidth}\leavevmode\epsfbox{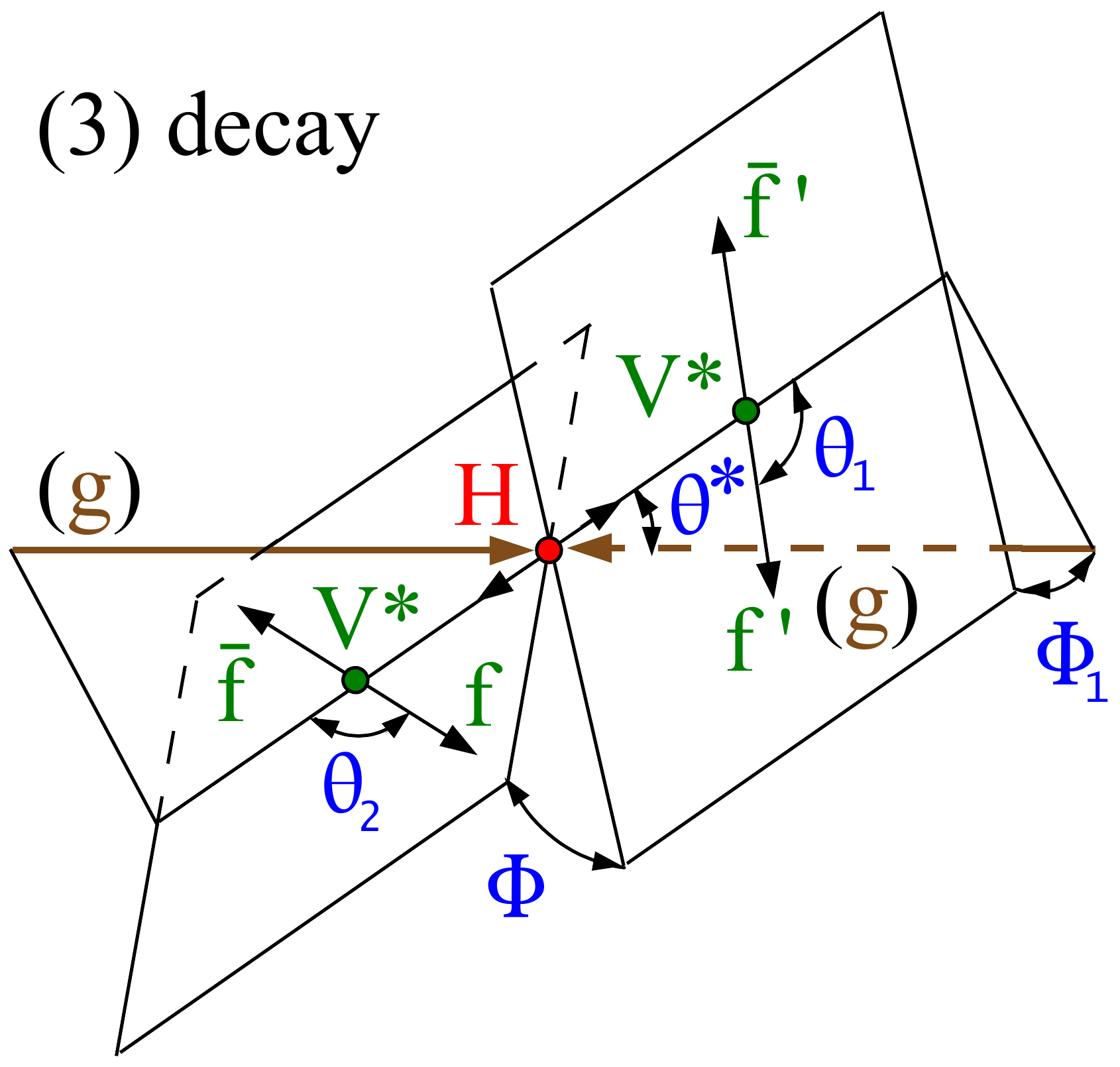}
}
\captionsetup{justification=centerlast}
\caption{
Illustrations of the \Hboson kinematic observables in $pp$, $e^+e^-$, or $\mu^+\mu^-$ collision in
(1) the VBF process: $e^+e^-(q{q^\prime})\to e^+e^-(q{q^\prime})H \to e^+e^-(q{q^\prime}) b\bar{b}$;
(2) the $VH$ process: $e^+e^-(q\bar{q})\to V^*\to V^*H\to f\bar{f}b\bar{b}$;
and (3) the decay: gg\,$\to H\to V^*V^*\to 4f$.
Five angles fully characterize the orientation of the production and decay chain and are defined 
in the suitable rest frames.  
The diagrams are adopted from Refs.~\cite{Gao:2010qx,Anderson:2013afp}.
}
\label{fig:decay}
\end{figure}

The LHC could be considered a gluon collider, as the dominant 
\Hboson production mechanism is the gluon fusion gg~$\to H$ process. 
Many aspects discussed in Section~\ref{sect:photon} in application to the photon couplings apply here as well. 
The coupling of the \Hboson to two massless gluons is generated by the loops of any massive particles with
color charge, which are quarks in the SM. In BSM, new heavy states, either fermions or bosons, could 
contribute to the loop and generate $CP$ violation. While there is a sizable decay rate $H\to$~gg, study of
the gluon polarization is difficult, and within the hadron collider environment this decay mode is hard to distinguish 
from the dominant QCD background. However, study of the gluon fusion process in the scattering topology 
of the \Hboson production in association with two hadronic jets allows access to the $CP$ property of the 
$H$gg vertex~\cite{Hankele:2006ma}. This VBF topology is illustrated in the first diagram of Fig.~\ref{fig:decay},
where $V^*=$\,g. 

Similarly to the photon couplings, the gluon fusion process can be characterized by the two 
couplings $CP$-even $a_2^{H\rm gg}$ and $CP$-odd $a_3^{H\rm gg}$, which absorb both 
SM and heavy BSM particles in the loop. 
While in the EFT approach, these contributions could be disentangled in a global fit of multiple
processes, this is not possible with the gluon fusion process alone. Therefore, we parameterize the 
$CP$ violation effects with a single parameter $f_{CP}^{H\rm gg}$. In order to isolate $CP$-sensitive
effects, no constraint on the process rate is applied, which is proportional to 
$|a_2^{H\rm gg}|^2+|a_3^{H\rm gg}|^2$.

The Snowmass-2013 projection~\cite{Dawson:2013bba} was based on the study of the gg~$\to H$ 
process with $H\to4\ell$~\cite{Anderson:2013afp}. Since then, this approach was successfully applied 
on LHC~\cite{Sirunyan:2021fpv,Collaboration:2022mlq,ATLAS:2021pkb} using 140\,fb$^{-1}$ of data,
with the results in good agreement with the above expectation. For example, the constraint $f_{CP}^{H\rm gg}<0.26$ 
at 68\% C.L. is expected with $H\to\tau\tau$ and $4\ell$ combined~\cite{Collaboration:2022mlq}, 
which would scale to $f_{CP}^{H\rm gg}<0.011$ with 3000\,fb$^{-1}$, as shown in supplemental materials~\cite{CMS-HIG-20-007},
and $f_{CP}^{H\rm gg}<0.12$ with 300\,fb$^{-1}$, where the improvement with respect to the Snowmass-2013 
projection is due to an additional \Hboson decay channel analyzed.
Somewhat more conservative results are projected in Ref.~\cite{Gritsan:2020pib}. 
Therefore, in Table~\ref{table-cpscenarios},
we estimate that $f_{CP}^{H\rm gg}<0.011$ could be achieved with 3,000\,fb$^{-1}$,
but note that further improvement is very likely, both from the inclusion of multiple decay channels and
from improvements in experimental analyses. 

\subsection{The $H\gamma\gamma$ and $HZ\gamma$ couplings at a hadron collider}
\label{sect:HAA}

While the $H\to\gamma\gamma$ decay was one of the two primary \Hboson discovery channels, this decay
process does not allow access to $CP$ structure of the photon couplings, as discussed in Section~\ref{sect:photon}.
Similarly, the $H\to Z\gamma$ decay does not provide access to $CP$ structure of the $HZ\gamma$ vertex
when the couplings in Eq.~(\ref{eq:fullampl-spin0}) are real. Complex couplings could be generated with 
light particles in the loop, for example, and could generate forward-backward asymmetry in the polar angle, 
as shown in Ref.~\cite{Anderson:2013afp}, but we do not consider such a possibility here. 
Another approach, suggested in Ref.~\cite{Chen:2014ona}, relies on complex phases generated through 
different Breit-Wigner propagators in the interference of the $\gamma^*$ and $Z$ contributions in the decay 
$H\to \ell^+\ell^-\gamma$ process. However, this approach is not necessarily better than $H\to4\ell$,
which includes $H\to \ell^+\ell^-(\gamma^*/Z)$, and would require further feasibility studies. 

However, several other topologies may allow access to the $CP$ structure of the $H\gamma\gamma$ 
and $HZ\gamma$ couplings. These include
\begin{enumerate}
\item the VBF process: $qq^\prime\to qq^\prime(\gamma^*\gamma^*/Z\gamma^*)\to qq^\prime H$, 
\item the $VH$ process: (a) $q\bar{q}\to\gamma^*/Z\to\gamma^*H/ZH\to (2f)H$, ~~~(b) $q\bar{q}\to\gamma^*/Z\to\gamma H$,
\item the 4/3/2-body decays: (a) $H\to \gamma^*\gamma^*/Z\gamma^*\to 4f$, ~~~(b) $H\to \gamma\gamma^*/\gamma Z\to\gamma(2f)$, ~~~(c) $H\to\gamma\gamma$.
\end{enumerate}
It is important to note that the above processes do not appear in isolation and whenever $\gamma^*$ appears in 
the intermediate state, $Z^*$ appears as well, leading to interference. For example, the full analysis of the process 
$H\to \gamma^*\gamma^*/\gamma^*Z/ZZ\to 4f$ may allow access to $CP$ violation through interference
of the $CP$-odd $a_3^{HZ\gamma}$ term with the dominant $CP$-even $a_1^{HZZ}$ term
appearing at tree level, and this needs to be disentangled from the possible $a_3^{H\gamma\gamma}$
and $a_3^{HZZ}$ terms. Therefore, the full analysis of each process requires accounting for all contributions,
including the $HZZ$ couplings discussed in Section~\ref{sect:HVV}. 

The three LHC topologies which involve the $H\gamma\gamma$ and $HZ\gamma$ couplings are shown in Fig.~\ref{fig:decay},
where $V^*=\gamma^*$ or $Z$. The processes with onshell photons can also be represented by the diagrams (2) and (3)
in Fig.~\ref{fig:decay} with $V^*=\gamma$, but with no subsequent decay of the photon. However, these 
processes with onshell photons do not allow access to $CP$ effects without the measurement of
the onshell photon polarization, unless complex anomalous couplings are considered. 
Illustration of $CP$-sensitive effects with complex couplings appearing in the forward-backward asymmetry 
of the angular distributions can be found for $q\bar{q}\to Z^*\to\gamma H$ in Ref.~\cite{Davis:2021tiv} 
and $H\to \gamma Z\to\gamma(2f)$ in Ref.~\cite{Anderson:2013afp}.
An angle $\Phi$ identified in all three diagrams in Fig.~\ref{fig:decay} is an angle between the decay or
production planes defined by the four four-momenta and is the primary $CP$-odd observable in 
each process. However, a multivariate analysis of the full kinematic information leads to the most 
optimal amplitude analysis, which is sensitive to both squared and interference of the $CP$-odd
and $CP$-even terms. 

An attempt to study the $CP$ structure of the $H\gamma\gamma$ and $HZ\gamma$ couplings in the golden 
channel $H\to4\ell$ was performed at the LHC in Ref.~\cite{Khachatryan:2014kca}, where it became clear that
reaching an interesting level of sensitivity will require very high luminosity. The $H\gamma\gamma$ 
couplings in the $H\to4\ell$ decay were considered phenomenologically in Ref.~\cite{Chen:2014gka}. 
More recently, a joint analysis of the three processes VBF, $VH$, and decay $H\to4\ell$ 
was investigated in Ref.~\cite{Davis:2021tiv}, where it was shown that while 
the decays $H\to\gamma\gamma$ and $H\to Z\gamma$ are most sensitive to the overall strength
$|a_2^{H\gamma\gamma}|^2+|a_3^{H\gamma\gamma}|^2$ and $|a_2^{HZ\gamma}|^2+|a_3^{HZ\gamma}|^2$,
the decay $H\to4\ell$ process is most sensitive to study the tensor structure of the $H\gamma\gamma$ and $HZ\gamma$ couplings, 
relevant for $CP$ violation measurements. We should note that in all the above studies, the effective values of 
$a_2^{H\gamma\gamma}$ and $a_2^{HZ\gamma}$ which reproduce the SM rate of $H\to\gamma\gamma$ and $H\to Z\gamma$
decays were used to approximate the SM processes $H\to \gamma^*\gamma^*/Z\gamma^*\to 4\ell$, VBF, and $VH$.
While this prescription is not technically correct to represent the SM rate due to $q^2$ dependence of the couplings, 
this simulated value of $a_2$ is a good benchmark for the Snowmass exercise as it is used only as
a reference to estimate sensitivity to $a_3$. 
The results of the study in Ref.~\cite{Davis:2021tiv} are reinterpreted in terms of 
$f_{CP}^{H\gamma\gamma}$ and $f_{CP}^{HZ\gamma}$ in Appendix~\ref{app:A}
and are entered in Table~\ref{table-cpscenarios}. The full dataset of the HL-LHC will be at the 
boundary to start setting meaningful constraints on these parameters. 

Given the difficulty to set $CP$ constraints on the $H\gamma\gamma$ and $HZ\gamma$ couplings at HL-LHC, 
exploring other options might be useful. A possible study of $H\to\gamma\gamma\to4e$ with photon polarization 
in its conversion has been suggested in Ref.~\cite{Bishara:2013vya}, but this study suffers from a significant loss 
of statistical precision, and a more detailed study of experimental aspects, such as reconstruction of displaced 
and boosted $e^+e^-$ pairs with a small opening angle, may be required.


\subsection{The $HZZ$ and  $HWW$ couplings at a hadron collider}
\label{sect:HVV}

The $HZZ$ and $HWW$ couplings appear at tree level in the SM and the decays $H\to ZZ\to 4f$ and $H\to W^+W^-\to 4f$
provided rich kinematic information for studies of spin and $CP$ properties of the \Hboson in the early days after the \Hboson
discovery and have historically been studied extensively on LHC experiments~\cite{Chatrchyan:2012jja,
Aad:2013xqa,Chatrchyan:2013mxa,Khachatryan:2014kca,Aad:2015mxa,Khachatryan:2016tnr,
Aad:2016nal,Sirunyan:2017tqd,Aaboud:2017oem,Aaboud:2017vzb,Sirunyan:2019twz,
Sirunyan:2019nbs,Aad:2020mnm,Sirunyan:2021fpv,ATLAS:2021pkb,CMS:2022ley,Collaboration:2022mlq}. 
However, with the growing significance of the \Hboson electro-weak production (VBF and $VH$), 
the larger $q^2$ values tested lead to stronger constraints of $CP$ effects in these production modes. 
The three main topologies involving the $HZZ$ and $HWW$ couplings follow closely those in Section~\ref{sect:HAA}
and appear in Fig.~\ref{fig:decay}, with $V^*=Z$ or $W$, as
\begin{enumerate}
\item the VBF process: $qq^\prime\to qq^\prime(W^+W^-/ZZ)\to qq^\prime H$, 
\item the $VH$ process: (a) $q\bar{q}\to Z\to ZH\to (2f)H$, ~~~(b) $q\bar{q}^\prime\to W^\pm\to W^\pm H\to (2f)H$,
\item the 4-body decay: (a)  $H\to ZZ\to 4f$, ~~~(b) $H\to W^+W^-\to 4f$.
\end{enumerate}
All processes with the $Z$ boson interfere with the same processes involving $\gamma^*$ in its place, 
as listed in Section~\ref{sect:HAA}. We note that the process gg~$\to VH$ also receives attention due to the large 
gluon luminosity in proton collisions. This channel provides an interesting interplay of $\Hff$ and $HVV$ couplings, 
but does not have contribution from the $CP$-odd terms with $a_3^{HZZ}, a_3^{HZ\gamma},$ or 
$a_3^{H\gamma\gamma}$~\cite{Gritsan:2020pib}, and therefore is not suitable for studies of $CP$ violation in 
$HZZ, {HZ\gamma}$, or ${H\gamma\gamma}$ interactions. 

Even though the $HZZ$ and $HWW$ couplings can be easily analyzed separately in the $ZH$ vs. $WH$ production
with leptonic $Z$ or $W$ decay, or in $H\to ZZ$ vs. $H\to WW$ decays, it is essentially impossible to disentangle those 
in the VBF production, where all kinematic features are nearly identical. 
The tree-level couplings $HZZ$ and $HWW$ can be related through 
custodial symmetry, leading to $a_1^{HZZ}=a_1^{HWW}$. Within the precision of the \Hboson measurements, 
this relationship is not significantly affected by the recent tension in the $W$ mass measurements.
The anomalous $HZZ$ and $HWW$ couplings, 
such as $a_3^{HZZ}$ and $a_3^{HWW}$, could also be related through symmetry, such as $SU(2)\times U(1)$. 
For example, $a_3^{HWW}=a_3^{HZZ}\!\cdot\!\cos^2\theta_W$ if contributions 
of the $H\gamma\gamma$ and $HZ\gamma$ couplings are neglected. 
Most of the experimental studies on LHC
and projections of the feasibility studies have been performed under such or a similar relationship 
of the $HZZ$ and $HWW$ couplings. Therefore, in Table~\ref{table-cpscenarios} we estimate 
precision on $f_{CP}^{HVV}$ which represents $V=Z$ and $W$ combined.\footnote{Technically, $f_{CP}^{HVV}$
is defined for $H\to ZZ\to 2e2\mu$ with $\sigma_1^{HZZ}/\sigma_3^{HZZ}=6.54$ in Eq.~(\ref{eq:fractions}),
but the measurement relies on both $HZZ$ and $HWW$ couplings.}
Precision of the separate measurements would be less, but similar. 
We should also note that since $a_1^{HZZ}$  and $a_1^{HWW}$ are generated at tree level in the SM, 
they are expected to be much larger than $a_3^{HZZ}$ and $a_3^{HWW}$, which appear at loop level, 
similar to the photon couplings. Therefore, the interesting values of $f_{CP}^{HVV}$ are much smaller 
than those for $f_{CP}^{H\rm gg}$, $f_{CP}^{H\gamma\gamma}$, and $f_{CP}^{HZ\gamma}$, 
as reflected in the last column of Table~\ref{table-cpscenarios}. 

The Snowmass-2013 projections~\cite{Dawson:2013bba} were split into the study of the $HVV$ 
couplings in three processes: $H\to4\ell$ decay, VBF production with $H\to\gamma\gamma$, and 
$VH$ production with $H\to b\bar{b}$~\cite{Anderson:2013afp}, where the most powerful channels 
were picked in each case. In the present study, we do not separate the channels and
consider the combined or best performance, assuming that the effective field-theoretic description
does not breakdown with the $q^2$ growth. Several experimental updates with 140\,fb$^{-1}$ 
of LHC data have appeared since then, some of the recent ones include 
Refs.~\cite{Sirunyan:2021fpv,Collaboration:2022mlq}, where the constraint $f_{CP}^{HVV}<8\times 10^{-5}$
is expected at 68\% C.L. from analysis of electroweak production information in the $H\to\tau\tau$ and $4\ell$ channels. 
The \Hboson physics projections at the HL-LHC and HE-LHC were revised in Ref.~\cite{Cepeda:2019klc},
where Fig.~38 indicates $f_{CP}^{HVV}<0.037$ from $H\to4\ell$ and Fig.~39 indicates 
$f_{CP}^{HVV}<1.8\times 10^{-4}$ at 95\% C.L. from production with $H\to4\ell$ at 3,000\,fb$^{-1}$
When a combined analysis of the $H\to\tau\tau$ and $4\ell$ channels is used, 
the expectation is $f_{CP}^{HVV}<2.5\times 10^{-6}$ at 68\% C.L. at 3,000\,fb$^{-1}$,
as shown in supplemental materials~\cite{CMS-HIG-20-007}.
We use these studies to indicate that $f_{CP}^{HVV}<2.5\times 10^{-6}$ at 3,000\,fb$^{-1}$
and $f_{CP}^{HVV}<4.0\times 10^{-5}$ at 300\,fb$^{-1}$ are achievable at 68\% C.L.
Further improvements are likely from the inclusion of multiple decay channels and
from improvements in experimental analyses. 


\subsection{The $Ht\bar{t}$ coupling at a hadron collider}
\label{sect:Htt}

The $CP$ structure of the \Hboson couplings to fermions is particularly interesting because 
both $CP$-even and $CP$-odd components can appear at tree level, and therefore the 
$f_{CP}^{\Hff}$ values do not necessarily need to be very small.
(This is in contrast to $f_{CP}^{HVV}$, for example.)
One could get access to the ${\Hff}$ interactions through loops appearing in the 
${H\rm gg}$, $H\gamma\gamma$, and $HZ\gamma$, but we treat those separately 
in Sections~\ref{sect:ggH} and \ref{sect:HAA} because
one cannot disentangle loop contributions without a global analysis. 
A measurement of the $CP$ structure of the \Hboson couplings to the first- and 
second-family fermions is essentially impossible at a hadron collider, as there are no channels
where polarization measurements could be performed. A $CP$ measurement of the $H b\bar{b}$
vertex is also essentially impossible, as neither $H\to b\bar{b}$ decay nor $b\bar{b}H$ production
allows access to $CP$~\cite{Gritsan:2016hjl} without the polarization measurements of the $b$ quarks,
which, if attempted in their decays, suffer from significant loss of statistical precision. 
This leaves only the $Ht\bar{t}$ and $H\tau\tau$
couplings with $CP$ structure that can be measured at a hadron collider. 

\begin{figure}[t!]
\centerline{
\setlength{\epsfxsize}{0.31\linewidth}\leavevmode\epsfbox{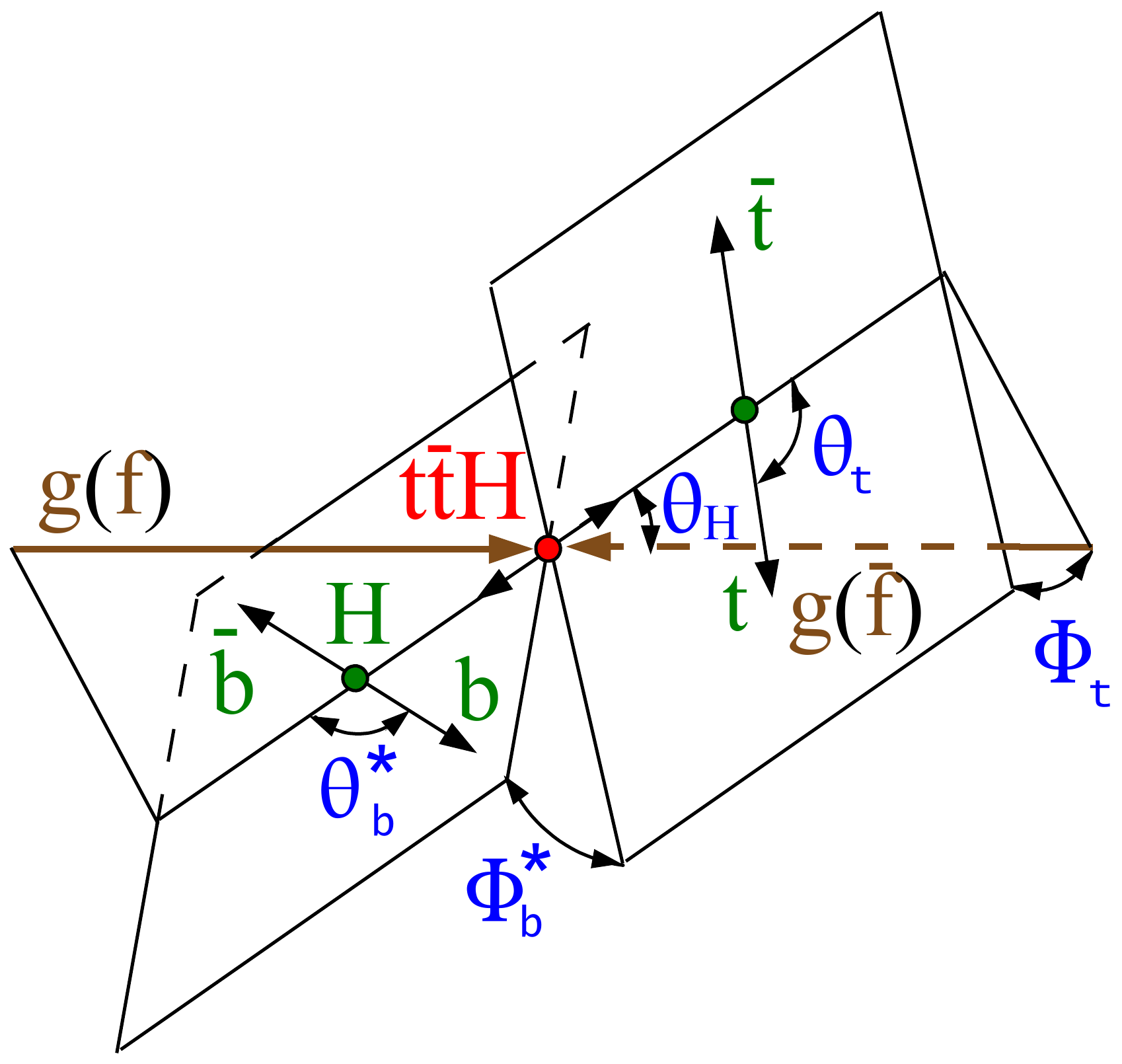}
\setlength{\epsfxsize}{0.31\linewidth}\leavevmode\epsfbox{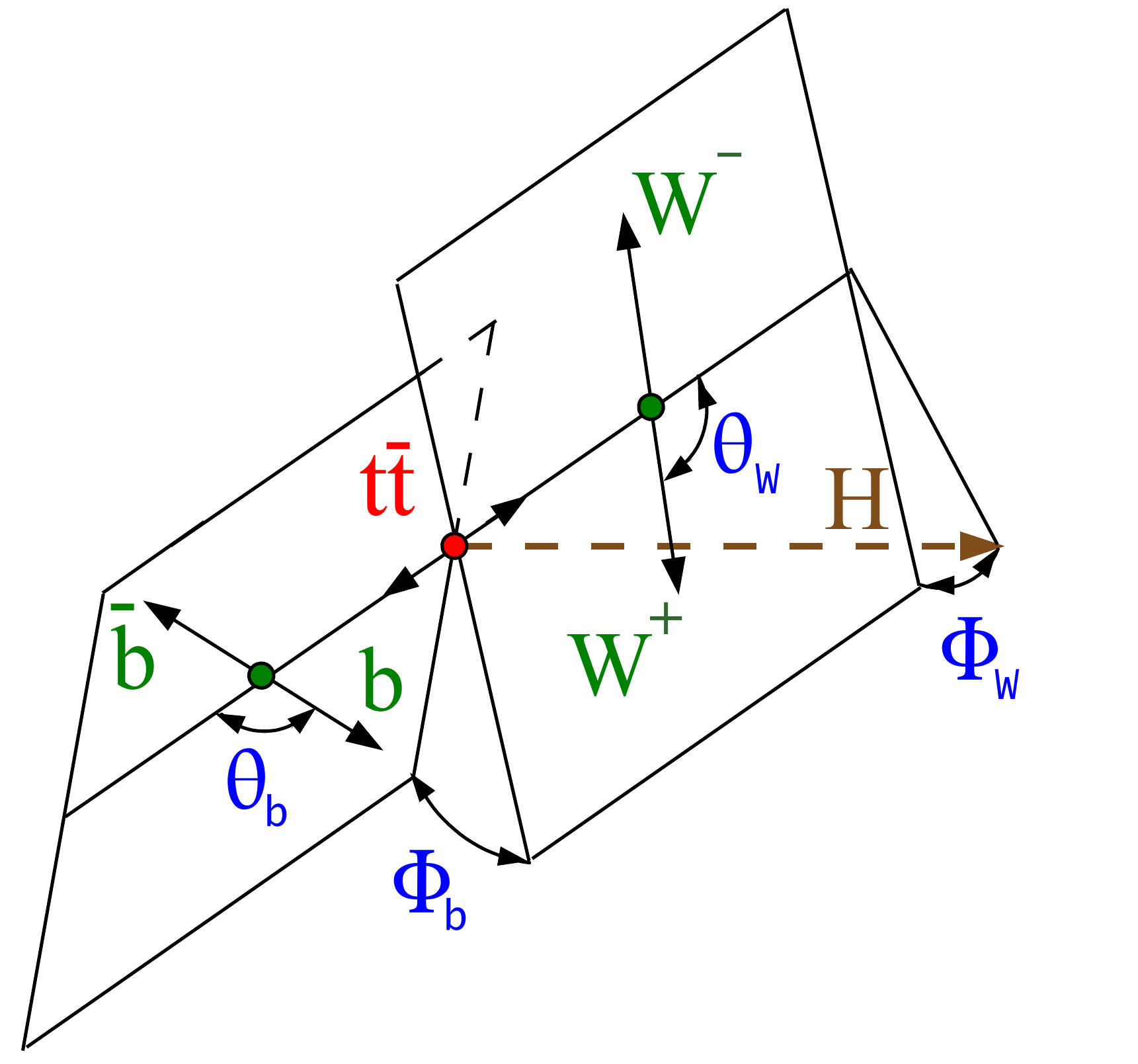}
\setlength{\epsfxsize}{0.31\linewidth}\leavevmode\epsfbox{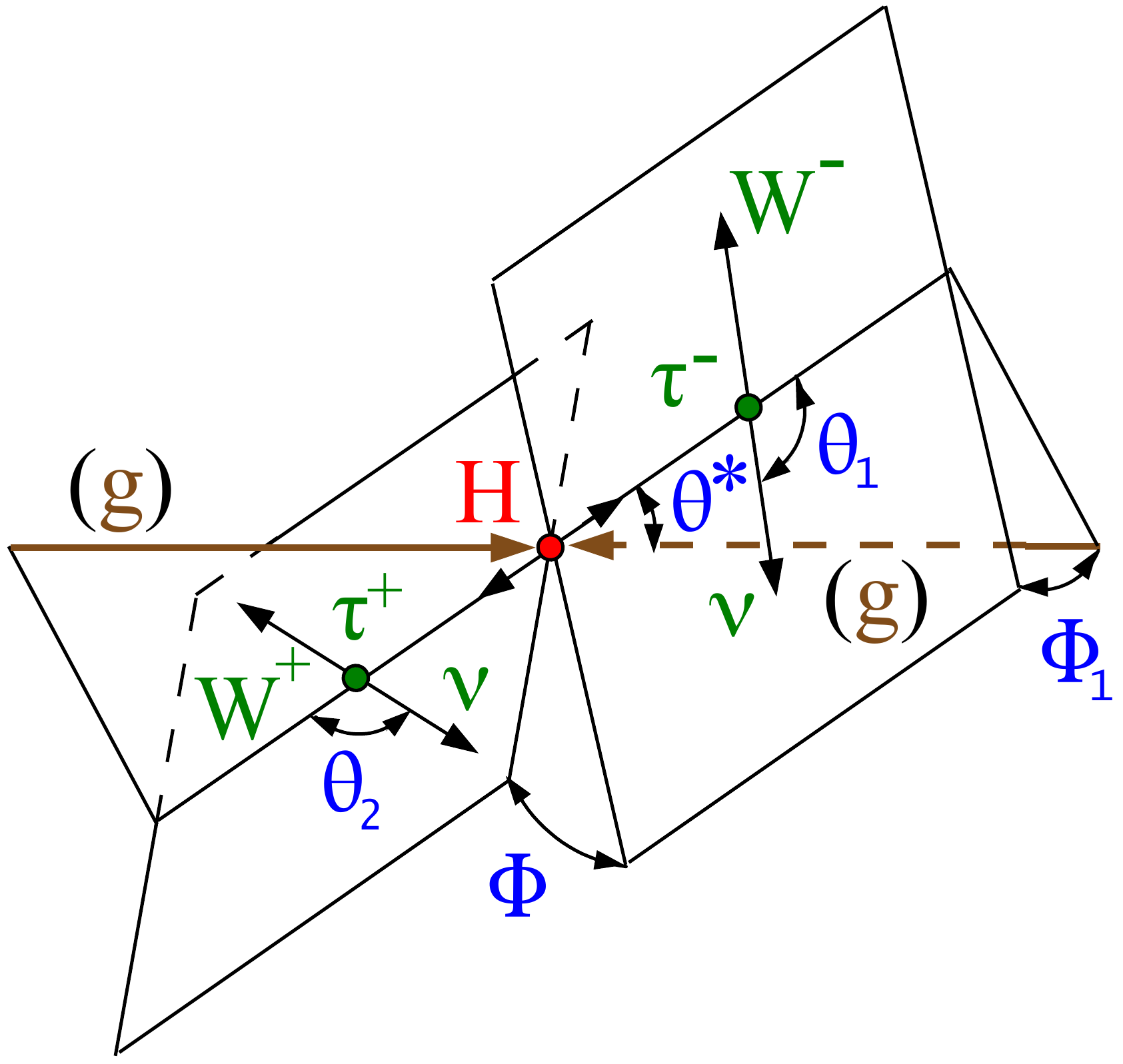}
}
\captionsetup{justification=centerlast}
\caption{
Illustrations of the \Hboson kinematic observables in $pp$, $e^+e^-$, or $\mu^+\mu^-$ collisions 
in (left and middle) the $t\bar{t}H$ process with sequential decay; 
(right) the $H\to \tau^+\tau^-$ decay process. The subsequent $W$ decay is not shown. 
The $b$ and $W$ pairing in the $t\bar{t}$ decays is switched to enhance visibility 
of $CP$ effects in individual angular observables. 
The diagrams are adopted from Ref.~\cite{Gritsan:2016hjl}.
}
\label{fig:Hff}
\end{figure}

The $t\bar{t}H$ production process has received the primary attention on LHC as the channel 
to study $CP$ the structure of the $Ht\bar{t}$ 
coupling~\cite{Gunion:1996xu,Demartin:2014fia,Buckley:2015vsa,Gritsan:2016hjl,Martini:2021uey,Barman:2021yfh,Barman:2022pip},
while the $tqH$ and $tWH$ processes also allow access to $CP$ in this coupling.
The cross sections of the latter channels are smaller, but they feature interference of the 
$Ht\bar{t}$ and $HVV$ couplings, which help in resolving the sign ambiguity in the relative phase,
and a joint analysis of all these channels is often required due to cross-feed of events in analysis of the data. 
There is also a proposal to access $CP$-violating effects in $Ht\bar{t}$ couplings through loop
effects in $t\bar{t}$ production~\cite{Martini:2021uey}, but the precision of such constraints does not
alter our conclusion drawn from channels with associated \Hboson production. 
There is rich kinematic information in the sequential decay of the particles produced in 
association with the \Hboson in the $t\bar{t}H$, $tqH$, and $tWH$ processes,
as indicated in the two diagrams in Fig.~\ref{fig:Hff}. However, most information 
is sensitive to the square of the $CP$-odd and $CP$-even amplitudes. 

It is also possible to construct $CP$-odd observables that are sensitive to the interference term in the amplitude 
by exploring the $t\bar{t}$ spin correlations. These spin correlations can be traced back from the decay products 
of $t$ and $\bar{t}$, since the top-quark lifetime~($\sim 10^{-25}~\mathrm{s}$~\cite{ParticleDataGroup:2020ssz}) 
is much shorter than the time required for spin decorrelation effects 
to actualize~($\sim 10^{-21}~\mathrm{s}$)~\cite{Mahlon:2010gw}. 
Thus, $CP$-odd observables can be constructed from antisymmetric tensor products of the four-momenta 
of the top, anti-top, and their respective decay products $i$ and $k$, 
$\epsilon(p_{t}, p_{\bar{t}}, p_{i}, p_{k}) \equiv \epsilon_{\mu \nu \rho\sigma}p_{t}^{\mu}p_{\bar{t}}^{\nu}p_{i}^{\rho}p_{k}^{\sigma}$. 
In the $t\bar{t}$ rest frame, this antisymmetric tensor product can be simplified to 
$\vec{p}_{t}\cdot\left(\vec{p}_{i} \times \vec{p}_{k} \right)$, which can be used to define genuine 
$CP$-sensitive azimuthal angle differences~\cite{Goncalves:2018agy,Goncalves:2021dcu}. 
Correlations between two decay products, one from $t$ and the other from $\bar{t}$, scale with the spin 
analyzing power ($\beta_{i}$) associated with the decay product~\cite{Bernreuther:2010ny}. 
Charged leptons and down-type quarks exhibit the highest spin correlations $|\beta_{i}| = 1$, 
followed by bottom quarks and W bosons with $|\beta_{i}| = 0.4$, and neutrinos and up-type quarks 
with $|\beta_{i}| = 0.3$. Therefore, one would require access to the flavors of the fermions and 
anti-fermions in the subsequent $t/W^+$ and $\bar{t}/W^-$decays to probe the interference. 
This is possible in the leptonic decays of both $W$'s in the $t\bar{t}$ decay, but statistical precision 
in this fully leptonic channel is significantly weaker than in the semi-leptonic and fully-hadronic channels. 

Both the CMS and ATLAS experiments~\cite{Sirunyan:2020sum,ATLAS:2020ior,Sirunyan:2021fpv,
CMS-PAS-HIG-21-006,ATLAS-CONF-2022-016}
have performed an amplitude analysis of the $CP$-even and $CP$-odd components of
the $Ht\bar{t}$ coupling analyzing both $t\bar{t}H$ and $tH$ processes. One of the dominant 
\Hboson decay channels is $H\to\gamma\gamma$, but other decays can also make significant 
contributions. Only semi-leptonic and fully-hadronic top decays in the $t\bar{t}H$ channels 
have been used, therefore limiting $CP$ analysis to the square of the amplitudes. 
With about 140\,fb$^{-1}$ of LHC data, a single experiment obtained an expected sensitivity
of $f_{CP}^{Ht\bar{t}}<0.5$ at 68\% C.L. with $H\to\gamma\gamma$~\cite{Sirunyan:2020sum} and 
of $f_{CP}^{Ht\bar{t}}<0.35$ in combination with the multi-lepton \Hboson decays~\cite{CMS-PAS-HIG-21-006}.
Phenomenological studies indicate
$f_{CP}^{Ht\bar{t}}<0.4$ at 300\,fb$^{-1}$ in Ref.~\cite{Gritsan:2016hjl} using the $H\to\gamma\gamma$ channel. 
Exploring the same \Hboson final state together with combined searches in the semi-leptonic, di-leptonic, and 
fully-hadronic top quark pair decays, the study in Ref.~\cite{Barman:2021yfh} indicates a sensitivity of 
$f_{CP}^{Ht\bar{t}}<0.05$ at the HL-LHC with 3,000\,fb$^{-1}$. Despite larger rates, searches in the 
$pp \to t\bar{t}(H \to b\bar{b})$ channel leads to typically weaker projections due to an imposing QCD background, 
which is also marred by substantial systematic uncertainties~\cite{Goncalves:2018agy,Goncalves:2021dcu}. 
The diphoton channel stands at a vantage point due to controlled backgrounds facilitated by data-driven side-bands. 
Inclusive and differential \Hboson measurements were investigated in Ref.~\cite{Bahl:2020wee}
and full kinematic information was investigated phenomenologically in Ref.~\cite{Bahl:2021dnc}, 
leading to similar expectations. 
We enter $f_{CP}^{Ht\bar{t}}<0.24$ at 300\,fb$^{-1}$ and $<0.05$ at 3,000\,fb$^{-1}$
in Table~\ref{table-cpscenarios} and note that further improvements are expected
from analysis of other \Hboson decay channels.


\subsection{The $H\to \tau^+\tau^-$ process at a hadron collider}
\label{sect:HTT}

The $H\to \tau^+\tau^-$ decay is an excellent probe of spin correlation in the sequential decay of the two taus. 
For example, an angle between the two decay planes indicated in Fig.~\ref{fig:Hff} is sensitive to $CP$ in the 
$H\tau\tau$ interaction. However, this angle cannot be measured directly due to missing neutrinos, and the
experimental challenge is to approximate it with available information. For example, the pion is preferably 
emitted in the direction of the $\tau$ spin in the $\tau$ rest frame, and additional information, such as the tau 
decay impact parameter, help to reconstruct $CP$-sensitive observables. 

At the time of the Snowmass-2013 studies~\cite{Dawson:2013bba}, it was believed that this reconstruction 
would be challenging, though possible, in the hadron collider environment~\cite{Berge:2011ij}. 
Most studies were focussed on the cleaner $e^+e^-$ collider environment, discussed in Section~\ref{sect:eeHTT}. 
A study in Ref.~\cite{Harnik:2013aja} using an optimal observable based on the internal substructure of
$\tau^\pm\to\pi^\pm\pi^0\nu$ indicated sensitivity to $f_{CP}^{H\tau\tau}<0.04$ at 3,000\,fb$^{-1}$ integrated luminosity of HL-LHC. 
However, it was found in Ref.~\cite{Askew:2015mda} that detector effects would be more important than originally suggested. 
A realistic study by the ATLAS collaboration~\cite{ATL-PHYS-PUB-2019-008} was based on analysis of
$\tau^\pm\to\pi^\pm\pi^0\nu$ and indicated that at HL-LHC the statistical precision 
on $f_{CP}^{H\tau\tau}$ would range between 0.10 and 0.30, 
depending on the precision of the $\pi^0$ reconstruction.
Finally, a very detailed study of multiple $\tau$ decay channels by the CMS experiment~\cite{CMS:2021sdq} 
achieved an expected precision of $f_{CP}^{H\tau\tau}<0.13$ at 68\% C.L. with about 140\,fb$^{-1}$. 
A similar recent study by ATLAS~\cite{ATLAS-CONF-2022-032} leads to $f_{CP}^{H\tau\tau}<0.22$ at 68\% C.L. 
The CMS experiment provided projection to 3,000\,fb$^{-1}$ as supplemental materials~\cite{CMS-HIG-20-006}
to Ref.~\cite{CMS:2021sdq}, from which we expect $f_{CP}^{H\tau\tau}<0.07$ at 300\,fb$^{-1}$ 
and $<0.008$ at 3,000\,fb$^{-1}$, which are entered in Table~\ref{table-cpscenarios}. 


\section{Prospects of Higgs $CP$ measurements at an electron-positron collider}
\label{sect:lepton}

Many of the approaches to the \Hboson $CP$ measurements at an electron-positron collider are similar to those 
at a hadron collider, but with several notable features. First, the $e^+e^-$ collider environment is much cleaner, 
and therefore even with a smaller number of $H$ bosons produced, essentially every final state of its decay
may be used for tagging. Second, certain final states, most notably $\tau^+\tau^-$, could be reconstructed and 
analyzed for $CP$ structure with better efficiency. Third, the fixed initial-state energy in the $e^+e^-\to V^*\to VH$ 
production allows control over the $q^2$ of the initial $V^*$. Similarly, possible polarization of the colliding beams 
may give additional control in polarization measurements. 

The $CP$ structure of the \Hboson couplings to gluons cannot be easily measured 
at a lepton collider, because the decay to two gluons does not allow easy access to gluon polarization. 
On the other hand, most other processes could be studied at an $e^+e^-$ collider, especially with the beam
energy above the $t\bar{t}H$ threshold. 


\subsection{The $VH$ process at an electron-positron collider}
\label{sect:eeVH}

The $e^+e^-\to ZH/\gamma^*H\to (2f)H$ process is the dominant SM process at lower energies 
with cross section of about $240/129/57/13$\,fb at $\sqrt{s}=250/350/500/1,000$\,GeV. 
Full angular analysis of the final state allows access to $CP$ information. 
Similarly to the hadron collider, the process $e^+e^-\to \gamma H$ is possible to study, 
but does not allow access to $CP$ properties from the angular analysis. This channel
has been used at LEP to set constraints on the \Hboson production with possible
anomalous $HZ\gamma$ and $H\gamma\gamma$ couplings. 

An early feasibility study of spin-parity determination and analysis of the $HZZ$ and $HZ\gamma$ coupling 
tensor structure in the $VH$ process at an $e^+e^-$ collider was performed as part of the TESLA 
design~\cite{ECFADESYLCPhysicsWorkingGroup:2001igx}
based on 300 fb$^{-1}$ at a centre-of-mass energy of 500 GeV and $m_H=120$ GeV.
The Snowmass-2013 studies~\cite{Dawson:2013bba} relied on Ref.~\cite{Anderson:2013afp}, 
which compared the expected performance of an $e^+e^-$ collider and the LHC, 
with the $H\to b\bar{b}$ and $Z\to\ell\ell$ decays used in the former case. 
Precision on the fraction of the $CP$-odd cross-section contribution of about 0.03 was obtained 
across the four energy and luminosity scenarios.
The significant reduction in the $f_{CP}^{HVV}$ uncertainties
with energy is due to the increase of the $q^2$ of intermediate $Z$, and therefore higher relative contribution 
of the higher-dimension operators to the production cross section, where it is assumed that no strong momentum 
dependence of couplings occurs at these energies.  
A recent update of these studies in Appendix~\ref{app:B} produced expected constraints with 
the same luminosity scenarios and with an order of magnitude higher integrated luminosity,
which are consistent within a factor of two with the more recent recommendations for Snowmass-2022 
studies, as outlined in Ref.~\cite{Dawson:2022zbb}. A study of beam polarization effects is also 
included in Appendix~\ref{app:B}.

There were no separate studies of the precision on $f_{CP}^{HZ\gamma}$ or $f_{CP}^{H\gamma\gamma}$ 
at the time of the Snowmass-2013 studies~\cite{Dawson:2013bba}, 
but a recent update of these studies in Appendix~\ref{app:B} indicate that it is not feasible
to constrain $f_{CP}^{HZ\gamma}$ or $f_{CP}^{H\gamma\gamma}$ at an $e^+e^-$ collider with 
parameters listed in Table~\ref{table-cpscenarios}. 
Only the $f_{CP}^{Z\gamma}$ parameter at $\mathrm{E}=1$\,TeV and  ${\cal L}=10$\,ab$^{-1}$
allows a non-trivial constraint at 68\% C.L., as indicated in the last column of Table~\ref{table-ee-revised}
in Appendix~\ref{app:B}, but still with essentially 100\% uncertainties. 
Of course, should there be an anomalously large $H{\gamma\gamma}$ or $H{Z\gamma}$ coupling,
much larger than expected from loop effects in the SM,
one could isolate $CP$-odd contributions with a relatively high precision, but such a scenario 
is excluded by the rates of the $H\to{\gamma\gamma}$ and $H\to{Z\gamma}$ processes.

The $CP$-odd $HZZ$ and $HZ\gamma$ couplings have been investigated with 5.6\,ab$^{-1}$ at 240\,GeV
with $Z\to\mu^+\mu^-$ and $H\to b\bar{b}, c\bar{c}$, gg in Ref.~\cite{Sha:2022bkt}, which follows closely 
similar earlier studies in Refs.~\cite{Craig:2015wwr,Beneke:2014sba}.
The expected constraint on the $CP$-odd $HZ\gamma$ coupling $a_3^{Z\gamma}$ 
is about a factor of six larger than the SM $a_2^{Z\gamma}$ expectation~\cite{Davis:2021tiv},
which indicates that it is hard to constrain photon couplings in this process, in agreement with the conclusion 
reached above about $f_{CP}^{HZ\gamma}$ or $f_{CP}^{H\gamma\gamma}$. 

The expected constraint on the $CP$-odd $HZZ$ coupling $a_3^{ZZ}$, which is equivalent 
to $\tilde{c}_{zz}$ used in Ref.~\cite{Sha:2022bkt}, can be translated to the expectation 
$f_{CP}^{HVV}<3.7\!\cdot\!10^{-4}$ at 68\% C.L. assuming that this limit scales 
linearly with luminosity (or square root of luminosity for the coupling). This appears to be 
similar to the Snowmass-2013 expectation and in agreement with the constraint 
obtained in Appendix~\ref{app:B}, $f_{CP}^{HVV}<3.4\cdot10^{-4}$.
The expected constraints obtained at three
other energy and luminosity scenarios in Appendix~\ref{app:B} are nearly identical to those
obtained in the Snowmass-2013 studies~\cite{Dawson:2013bba, Anderson:2013afp} and we
keep those unchanged. We also consider an order of magnitude higher luminosity in Appendix~\ref{app:B}.

It has been pointed out in Ref.~\cite{Ogawa:2017bmg} that in addition to the process $e^+e^-\to ZH$ 
with $Z\to\ell\ell$ ($7\%$), one can use the $Z\to q\bar{q}$ ($70\%$) final states. While both 
reconstruction and backgrounds will be somewhat more challenging with hadronic jets, the larger
branching fraction will make the hadronic channel dominate. We have not included results from
Ref.~\cite{Ogawa:2017bmg} in our projections in Table~\ref{table-cpscenarios}, because the 
comparison to other projections is not fully resolved, but we point out that in the most optimistic scenario, the final state
$Z\to q\bar{q}$ provides an order of magnitude increase in the available data, and the expected results
would correspond to a ten times larger luminosity scenario in Appendix~\ref{app:B}. 
Therefore, in the summary Table~\ref{table-cpscenarios}, we enter results from Table~\ref{table-ee-revised}
corresponding to the $Z\to\ell\ell$ channel, but with a ten times larger luminosity. 
When it comes to a comparison of HL-LHC and $e^+e^-$ projections, 
where typically only one \Hboson decay channel is analyzed at HL-LHC (e.g. $H\to\tau^+\tau^-$),
the inclusion of other channels (e.g. $H\to b\bar{b}, \gamma\gamma, W^+W^-, ZZ$)
will increase the available data by a factor of 5 or so. 


\subsection{The VBF process at an electron-positron collider}
\label{sect:eeVBF}

VBF production with charged boson fusion $e^+e^-\to \nu_e\bar\nu_e(W^+W^-)\to \nu_e\bar\nu_e H$
is the dominant SM process at higher energies with cross section of $21/34/72/210$\,fb at $\sqrt{s}=250/350/500/1,000$\,GeV. 
However, there is essentially no kinematic information to analyze in the final state with missing neutrinos,
with the exception of the momentum of the \Hboson, which provides sensitivity to the higher-dimension 
operators, but does not allow one to separate $CP$-odd and $CP$-even contributions. 
Therefore, this channel is useful to study $CP$ in the subsequent \Hboson decays, though the lack of 
a vertex from associated particles makes certain techniques less reliable, as discussed in application 
to $H\to \tau^+\tau^-$ for example. 

VBF production with neutral boson fusion $e^+e^-\to e^+e^-(ZZ/Z\gamma^*/\gamma^*\gamma^*)\to e^+e^- H$
cross section is smaller than of the main VBF channel with associated neutrinos, but is still sizable at higher energies
with the SM cross section of $0.7/3/7/21$\,fb at $\sqrt{s}=250/350/500/1,000$\,GeV. 
Full angular analysis allows access to $CP$ information. 
For example, an ongoing study of the $ZZ$-fusion process at 1.4 TeV CLIC and 1 TeV ILC
are mentioned in Ref.~\cite{Bozovic-Jelisavucic:2022ivd}, and the first preliminary estimates have
been shown at ICHEP-2022 conference~\cite{ICHEP-Vukasinovic}. However, the precision in this 
channel at these intermediate energies does not surpass that expected from 
the $e^+e^-\to ZH\to (2f)H$ process, and therefore Table~\ref{table-cpscenarios} is not affected. 
While there is no dedicated study of the $CP$-odd $H\gamma\gamma$ and $HZ\gamma$ interactions 
in VBF production $e^+e^-\to e^+e^-(ZZ/Z\gamma^*/\gamma^*\gamma^*)\to e^+e^- H$, 
an analogy has been drawn to the VBF process at LHC in Appendix~\ref{app:B}, which indicates that
it is unlikely that $f_{CP}^{HZ\gamma}$ or $f_{CP}^{H\gamma\gamma}$ could be constrained 
at an $e^+e^-$ collider. 


\subsection{The $t\bar{t}H$ process at an electron-positron collider}
\label{sect:eettH}

\Hboson production in association with top quarks $e^+e^-\to t\bar{t}H$ is the fourth production channel for energies 
above the threshold around 500\,GeV, with cross section of $0.27/2.0$\,fb at $\sqrt{s}=500/1,000$\,GeV. 
Many of the techniques used at the LHC in Section~\ref{sect:Htt} and at a muon collider in Section~\ref{sect:hemuon}
can be employed at an $e^+e^-$ collider, with the diagram in Fig.~\ref{fig:Hff} representating 
kinematic information in the process. 

A study of ${C\!P}$-odd contribution in the $Ht\bar{t}$ coupling has been studied in the context 
of ILC~\cite{Dawson:2013bba}. Cross-section dependence on the coupling has been employed
and an uncertainty on $f_{CP}^{Htt}$ of 0.08 (0.29) at 1,000 (500) GeV center-of-mass energy has been estimated. 
A beam polarization of $(+0.2,-0.8)$~\cite{Price:2014oca} and $(+0.3,-0.8)$ is assumed at 1,000 and 500 GeV, respectively.
A more recent study indicates sensitivity to $f_{CP}^{Htt}$ of about 0.07 expected with 2,000\,fb$^{-1}$
and 1,400\,GeV~\cite{CLICdp:2018esa}, which employs a similar cross section dependence.
Interpretation of a cross-section deviation as an indication of ${C\!P}$-odd coupling contribution
is strongly model-dependent, but allows access to anomalous $Ht\bar{t}$ couplings.
An analysis of the full kinematic information could proceed in a manner similar to that 
employed at LHC and would benefit from the clean $e^+e^-$ collider environment with 
the beam energy constraints available. 
An improvement from using the differential information has been observed in Ref.~\cite{thesis-Zhang}.


\subsection{The $H\to \tau^+\tau^-$ and other decay processes at an electron-positron collider}
\label{sect:eeHTT}

At the time of the Snowmass-2013 exercise~\cite{Dawson:2013bba}, most $CP$ studies with $H\to \tau^+\tau^-$ 
were performed in a clean $e^+e^-$ environment, either in the decays $\tau\to\pi\pi\nu$~\cite{Desch:2003rw,Harnik:2013aja}, 
or in all final states~\cite{Berge:2012wm,Berge:2013jra}. All studies agree on a similar $f_{CP}^{H\tau\tau}$ 
precision of about $0.01$ for the typical scenarios in Table~\ref{table-cpscenarios}.
The precision becomes somewhat worse with an increased collider energy due to the reduced $ZH$ production 
cross-section, and this technique relies on the knowledge of the $Z$ vertex. 
A recent full simulation study of the ILC physics reach with 1,000\,fb$^{-1}$ at 250\,GeV indicates a very similar $f_{CP}^{H\tau\tau}$ 
precision of about $0.01$ with $\tau^\pm\to\pi^\pm\nu$ and $\tau^\pm\to\pi^\pm\pi^0\nu$\cite{Jeans:2018anq, Bozovic-Jelisavucic:2022ivd},
but additional $\tau$ lepton decays may bring an increase in sensitivity. 
We therefore leave the estimates in Table~\ref{table-cpscenarios} the 
same as in the Snowmass-2013 projection. Further improvements could be achieved using the 
lessons learned from the realistic analysis of the $H\to \tau^+\tau^-$ channel at LHC, as 
discussed in Section~\ref{sect:HTT}. 

Analysis of the other decay channels, most notably $H\to 4f$, could be performed at an $e^+e^-$ collider.
The clean collider environment would allow exploration of multiple final states, beyond just the golden 
channels with charged leptons used at LHC. However, as noted in Appendix~\ref{app:B}, the number
of produced $H\to ZZ\to 4f$ events at an $e^+e^-$ collider would be significantly smaller than the number 
of $H$ bosons produced in the golden clean channel $H\to 4\ell$ at a proton collider. 

The $e^+e^-$ collider may become a clean environment for studies of the other two-body final states 
involving hadronic decays, though their polarization measurements would be challenging. 
In Section~\ref{sect:Htt}, we have already mentioned that in the $H\to b\bar{b}$ decay,
a polarization measurement of the $b$-quark jets 
may allow measurement of the $CP$ structure of the $H b\bar{b}$ vertex.
Similarly, in the $H\to\mathrm{gg}$ decay, a polarization measurement of the gluon jets 
may allow a $CP$ measurement in the $H\mathrm{gg}$ coupling.
Feasibility studies of both measurements in realistic experimental environment need to be performed. 


\section{Prospects of Higgs $CP$ measurements at a lepton-hadron collider}
\label{sect:lepton-hadron}

The electron-proton collider allows production of the \Hboson in the VBF topology and top-associated production
in a relatively clean environment, without the complications arising from pile-up typical of hadron colliders~\cite{LHeC:2020van}.
Three main production mechanisms include VBF production of an \Hboson and a quark jet in association 
with a neutrino $\nu_e Hq$ or an electron $e^- Hq$, and single-top-quark associated production $\nu_e H \bar{t}$. 
For the beam energies of $E_p=7$\,TeV and $E_e=60$\,GeV, corresponding to $\sqrt{s}=1.3$\,TeV, the cross sections 
of the these processes are 0.197\,pb, 0.024\,pb, and 0.002\,pb~\cite{LHeC:2020van}. Therefore, with about 
1\,ab of integrated luminosity, more than 200,000 $H$ bosons can be produced. 

\subsection{The VBF process at a lepton-hadron collider}
\label{sect:epVBF}

In VBF production, $WW$ and $ZZ/Z\gamma^*/\gamma^*\gamma^*$ fusion can be easily separated with the signature 
$e^-p\to \nu_e Hq$ (charged current) or $e^-p\to e^- Hq$ (neutral current), respectively. 
Studies of these processes~\cite{Biswal:2012mp,Cakir:2013bxa,Sharma:2022epc}
have been performed without consideration of the photon couplings. However, similarly to the VBF process 
at a proton-proton collider discussed in Appendix~\ref{app:A}, we would expect the photon couplings 
$H\gamma\gamma$ and $HZ\gamma$ to have weak constraints. The most recent projections of constraints
on the $HWW$ and $HZZ$ $CP$-odd couplings have been reported in Ref.~\cite{Sharma:2022epc}. 

\subsection{The top-$H$ process at a lepton-hadron collider}
\label{sect:epTopH}

The top Yukawa coupling has been studied in the $e^-p\to \nu_e H \bar{t}$ process in Ref.~\cite{Coleppa:2017rgb}. 
There is an interference between the diagrams with the $HWW$ and $Hbb$/$Htt$ couplings. 
The cross section of this process exhibits a dependence on the faction of the $CP$-odd component in the $Htt$ 
interaction. This dependence has been explored in Ref.~\cite{Coleppa:2017rgb}. However, a study which 
considers variation of more than one parameter and using $CP$-sensitive observables would be desired. 
In Table~\ref{table-cpscenarios}, we indicate that a study of the $Htt$ and $HZZ/HWW$ interactions 
is possible at an $e^-p$ collider. It would be also interesting to investigate if there is a sensitivity 
to $CP$ structure in the $Hbb$ interaction. 


\section{Comparison to EDM measurements}
\label{sect:edm}

A dedicated Snowmass-2022 study of EDM measurements can be found in Ref.~\cite{Alarcon:2022ero}.
Asymmetry in the charge distribution along the particle's spin requires $T$ violation,
which is equivalent to $CP$ violation when invoking the $CPT$ theorem. 
The EDMs of atoms and molecules are sensitive to $CP$ violation in interactions of the \Hboson through loop effects. 
The SM values of these EDMs are beyond the current or planned experimental reach, which allows excellent null tests in the SM. 
The EDM constraints on $CP$-odd \Hboson couplings are typically stronger than those from direct \Hboson 
measurements~\cite{Pospelov:2005pr,Li:2010ax,Cirigliano:2016nyn,Cirigliano:2019vfc,Bahl:2022yrs}.
However, these constraints are set under an assumption that only one modification 
of the \Hboson coupling is present in the loop, and therefore no cancellation effect is allowed. 
With multiple $CP$-odd EFT operators, the EDM measurements set constraints on certain linear 
combinations of these operators, and direct constraints on the $CP$-odd operators in the \Hboson 
measurements provide complementary information. 

For example, the $H\gamma\gamma$, $HZ\gamma$, $HZZ$, as $H$gg induce EDMs through one-loop diagrams. 
Replacing the $HVV$ vertex with a fermion loop in these diagrams leads to two-loop graphs, through which 
the $Hff$ couplings can contribute. One can also analyze these interactions with simultaneous contributions of 
loops of SM particles together with BSM interactions and point-like $HVV$ interactions generated by heavy BSM states.  
At the same time, the second vertex of the \Hboson involves $Hu\bar{u}$, $Hd\bar{d}$, or $Hee$ interactions, 
where $CP$ violation could be introduced as well. Therefore, in general, EDMs receive contributions from 
a large number of $CP$-odd interactions, allowing for the possibility of cancellations. 
While this brings complications, the EDM measurements may also allow the only access to $CP$ violation 
in the $Hee$, $Hu\bar{u}$, and $Hd\bar{d}$ interactions. 
Resolving all constraints simultaneously will require direct measurements of the \Hboson couplings
in combination with EDM measurements. 
Moreover, it has not been experimentally established if the \Hboson couples to the first-family 
fermions. In case these couplings are absent or significantly suppressed, EDM measurements
provide no constraints of $CP$ violation in \Hboson interactions. 

As part of this Snowmass study, we examine EDM constraints on parameters in Table~\ref{table-cpscenarios}
and add the $Hee$, $Hu\bar{u}$, and $Hd\bar{d}$ couplings in Table~\ref{tab:edm}. 
These constraints from the present EDM measurements are obtained in Appendix~\ref{app:C}. 
In Table~\ref{tab:edm}, only one $CP$-odd $HX$ coupling is allowed to be present at a time.
As it can also be seen in Fig.~\ref{fig:edm} of Appendix~\ref{app:C}, constraints on individual couplings
$H\gamma\gamma$, $HZ\gamma$, $HZZ$ are essentially lost if two other couplings are allowed to be present,
but a big part of parameter space is still excluded from a correlated measurement. 
Most constraints on the parameters in Table~\ref{tab:edm} are dominated by the 
the current limit on electron EDM $d_e<1.1\times 10^{-29}$\,$e$\,cm~\cite{ACME:2018yjb}
from the ThO measurement,
while the $CP$-odd $H$gg, $Hu\bar{u}$, and $Hd\bar{d}$ couplings are constrained by
the neutron~\cite{Abel:2020pzs} and mercury~\cite{Graner:2016ses} EDMs.
The limit on the neutron EDM is $d_n<1.8\times 10^{-26}$\,$e$\,cm~\cite{Abel:2020pzs},
and the mercury EDM constraint is equivalent to a similar limit on $d_n$~\cite{Graner:2016ses}, 
for the couplings under consideration here.

\begin{table}[h!]
\center\small
\renewcommand{\arraystretch}{1.3}
\captionsetup{justification=centerlast}
\caption{Constraints on the parameter $\Big|\frac{f_{CP}^{HX}}{1-f_{CP}^{HX}}\Big|$ at $68\%$ C.L.\ 
from EDM measurements, assuming only one $CP$-odd $HX$ coupling is nonzero at a time.
Refer to Appendix~\ref{app:C} for more details. 
}
\vspace{-0.1cm}
\begin{tabular}{|ccccccccccc|}
	\hline\hline
$HX$ coupling & ~~${H\rm gg}$~~ & ~~${H\gamma\gamma}$~~ & ~~${HZ\gamma}$~~ & ~~${HZZ}$~~ & ~~${Ht\bar{t}}$~~ & ~~${Hu\bar{u}}$~~ & ~~${Hd\bar{d}}$~~ & ~~${H\tau\tau}$~~ & ~~${H\mu\mu}$~~ & ~~${Hee}$~~ 	
	\\
	\hline
${f_{CP}^{HX}}/{(1-f_{CP}^{HX})}<$
& $0.12 $&	$2.4\cdot 10^{-8}$&$	4.4\cdot 10^{-8}$&$	1.2\cdot 10^{-13}$&	$4.3\cdot 10^{-7}$& $0.72$ & $0.039$ &  $	2.2\cdot 10^{-2} $& $36$ & $1.1\cdot 10^{-6}$
\\
	\hline\hline
\end{tabular}
\label{tab:edm}
\end{table}

Over the next two decades, one could expect an order of magnitude increase in the precision 
of the electron EDM every 5-6 years, e.g. Fig.~5 in Ref.~\cite{Alarcon:2022ero}.
There is also a dramatic increase possible in the nucleon EDM measurements, 
e.g. Fig.~8 in Ref.~\cite{Alarcon:2022ero}.
There is a proposal to reach a precision on the proton EDM $d_p< 10^{-29}$\,$e$\,cm
using the proton storage ring within the next decade~\cite{Alexander:2022rmq}, which would 
be a big improvement over the current neutron EDM constraint. 
This may lead to an improvement by $10^3$ in constraints on $CP$-odd $H$gg, $Hu\bar{u}$, and $Hd\bar{d}$
couplings, and potentially to an improvement by $10^6$ in constraints on corresponding~$f_{CP}^{HX}$. 
We note that even under the assumption of one $CP$-odd contribution to EDM, the 
expected constraint on $f_{CP}^{H\rm gg}$ at the HL-LHC in Table~\ref{table-cpscenarios}
is stronger than the present EDM constraint in Table~\ref{tab:edm}. 
With the above potential improvement on the proton EDM using the proton storage ring, 
this will change. However, the HL-LHC constraints will be essential in order to analyze 
all $CP$-violating couplings in Table~\ref{tab:edm} 
without assumptions that only one $CP$-odd coupling is nonzero at a time. 



\section{Summary}
\label{sect:summary}

The search for $CP$ violation is an important research direction of future experiments in particle 
physics, as $CP$ violation is required for baryogengesis and cannot be sufficiently explained with present knowledge.
We have reviewed the status and prospects of the search for $CP$ violation in interactions 
of the Higgs boson ($H$) with either fermions or bosons at the current and future proposed facilities. 
The dedicated $CP$-sensitive measurements of the $H$ boson provide simple but 
reliable benchmarks that can serve as a guide to compare future facilities as part of the 
Particle Physics Community Planning Exercise (a.k.a. ``Snowmass"). 
These benchmarks are compared between proton, electron-positron, photon, and muon 
colliders in Table~\ref{table-cpscenarios}.

Hadron colliders provide essentially the full spectrum of possible measurements sensitive to $CP$ violation 
in the $H$ boson interactions accessible in the collider experiments, with the exception of interactions with 
light fermions, such as $H\mu\mu$.
The $CP$ structure of the $H$ boson couplings to gluons cannot be easily measured at a lepton collider, 
because the decay to two gluons does not allow easy access to gluon polarization. 
On the other hand, most other processes could be studied at an $e^+e^-$ collider, especially with the beam 
energy above the $t\bar{t}H$ threshold. Future $e^+e^-$ colliders are expected to provide comparable $CP$
sensitivity to HL-LHC in $Hff$ couplings, such as $Ht\bar{t}$ and $H\tau\tau$, and $HZZ/HWW$ couplings. 

A muon collider operating at the $H$ boson pole gives access to $CP$ structure of the $H\mu\mu$ 
vertex using the beam polarization. It is not possible to study the $CP$ structure in the decay because the muon 
polarization is not accessible. At a muon collider operating both at the $H$ boson pole and at higher energy, 
analysis of the $H$ boson decays is also possible. However, this analysis is similar to the studies performed 
at other facilities and depends critically on the number of the $H$ bosons produced and their purity.
A photon collider operating at the $H$ boson pole allows measurement of the $CP$ structure of the 
$H\gamma\gamma$ vertex using the beam polarization. Otherwise, the measurement of $CP$ in both 
$H\gamma\gamma$ and $HZ\gamma$ interactions is challenging and requires high statistics of
$H$ boson decays with virtual photons, which would require a production rate beyond that of the HL-LHC for 
sensitive measurements. 

Measurements of the electric dipole moments of atoms and molecules set stringent constraints 
on $CP$-violating interactions beyond the SM appearing in loop calculations. 
Assuming only one $CP$-odd $H$ boson coupling is nonzero at a time, EDM constraints can 
be interpreted as limits on $CP$ violation in the $H$ boson interactions, as shown in Table~\ref{tab:edm}. 
Such constraints are either tighter or expected to be tighter with EDM measurements projected 
in the next two decades when compared to $CP$ violation measurements in direct $H$ boson 
interactions at colliders. However, resolving all constraints simultaneously will require direct 
measurements of the $H$ boson couplings in combination with EDM measurements. Moreover, 
it has not been experimentally established whether the $H$ boson couples to the first-family fermions,
and if such couplings are absent or suppressed, EDM measurements provide no constraints of 
$CP$ violation in $H$ boson interactions.

In the end, we conclude that the various collider and low-energy experiments provide complementary 
$CP$-sensitive measurements of the $H$ boson interactions. The HL-LHC provides the widest spectrum 
of direct measurements in the $H$ boson interactions and is unique in measuring couplings to gluons, 
but it lacks the ability to set precise constraints on interactions with photons and muons. Such constraints 
may become possible with either photon or muon colliders operating at the $H$ boson pole. 
The electron-positron collider may allow constraints similar to HL-LHC in couplings to fermions 
and heavy weak bosons. Given the coverage provided by HL-LHC, we expect that a future $pp$
collider, such as FCC-hh or SPPC, will surpass HL-LHC and allow the furthest reach in $CP$-sensitive 
measurements of the $H$ boson interactions among the collider experiments. 

\bigskip

\noindent
{\bf Acknowledgments}:
We would like to thank all contributors of individual studies and participants of the ``Snowmass" community exercise. 
A.V.G., J.D., L.S.M.G., and S.K. thank the United States National Science Foundation for the financial support, under grant number PHY-2012584. 
R.K.B and D.G. thank the United States Department of Energy for the financial support, under grant number DE-SC0016013.
%


\clearpage

\appendix 

\section{Recent updates of the studies at a hadron collider}
\label{app:A}

{\it Contributed by Jeffrey Davis, Savvas Kyriacou, and Jeffrey Roskes.}
\\

In this Section, we update the feasibility study of the $CP$-odd $H\gamma\gamma$ and $HZ\gamma$ interactions 
at the HL-LHC, which is documented in Ref.~\cite{Davis:2021tiv}, in order to adopt the $f_{CP}^{HV\gamma}$ benchmark parameters 
introduced in Eq.~(\ref{eq:fCP}). 
As discussed in Sections~\ref{sect:photon} and~\ref{sect:HAA}, it is not possible to study the $CP$ structure of the
$H\gamma\gamma$ and $HZ\gamma$ couplings in the $H\to\gamma\gamma$ and $H\to Z\gamma$ decays. 
The rates of these decays put constraints on the quadrature sum of the $CP$-odd and $CP$-even couplings,
which can be parameterized, following the notation in Eq.~(\ref{eq:gHX}) and in Ref.~\cite{Davis:2021tiv}, as
\begin{eqnarray}
\label{eq:ratio-Vgamma}
{\rm g}^{\rm eff~2}_{HV\gamma}\equiv
\frac{\Gamma_{H\to V\gamma}}{\Gamma_{H\to V\gamma}^{\rm SM}} \simeq
\frac{1}{\left(a_{2}^{V\gamma, \rm SM}\right)^2}
\left[
 \left(a_{2}^{V\gamma, \rm SM} + a_{2}^{V\gamma}\right)^2
+ \left(a_{3}^{V\gamma}\right)^2
\right]
\,,
\end{eqnarray}
where $V=Z$ or $\gamma$ and $a_{2}^{\gamma\gamma, \rm SM}=0.00423$ and $a_{2}^{Z\gamma, \rm SM}=0.00675$ 
are the effective values of the point-like $CP$-even couplings generated by SM loops
with the $W$ boson and charged fermions. In this parameterization, the SM corresponds to 
$(a_{2}^{\gamma\gamma}, a_{3}^{\gamma\gamma})=(0,0)$ and $(a_{2}^{Z\gamma}, a_{3}^{Z\gamma})=(0,0)$. 

\begin{figure}[h!]
\includegraphics[width=0.4\textwidth]{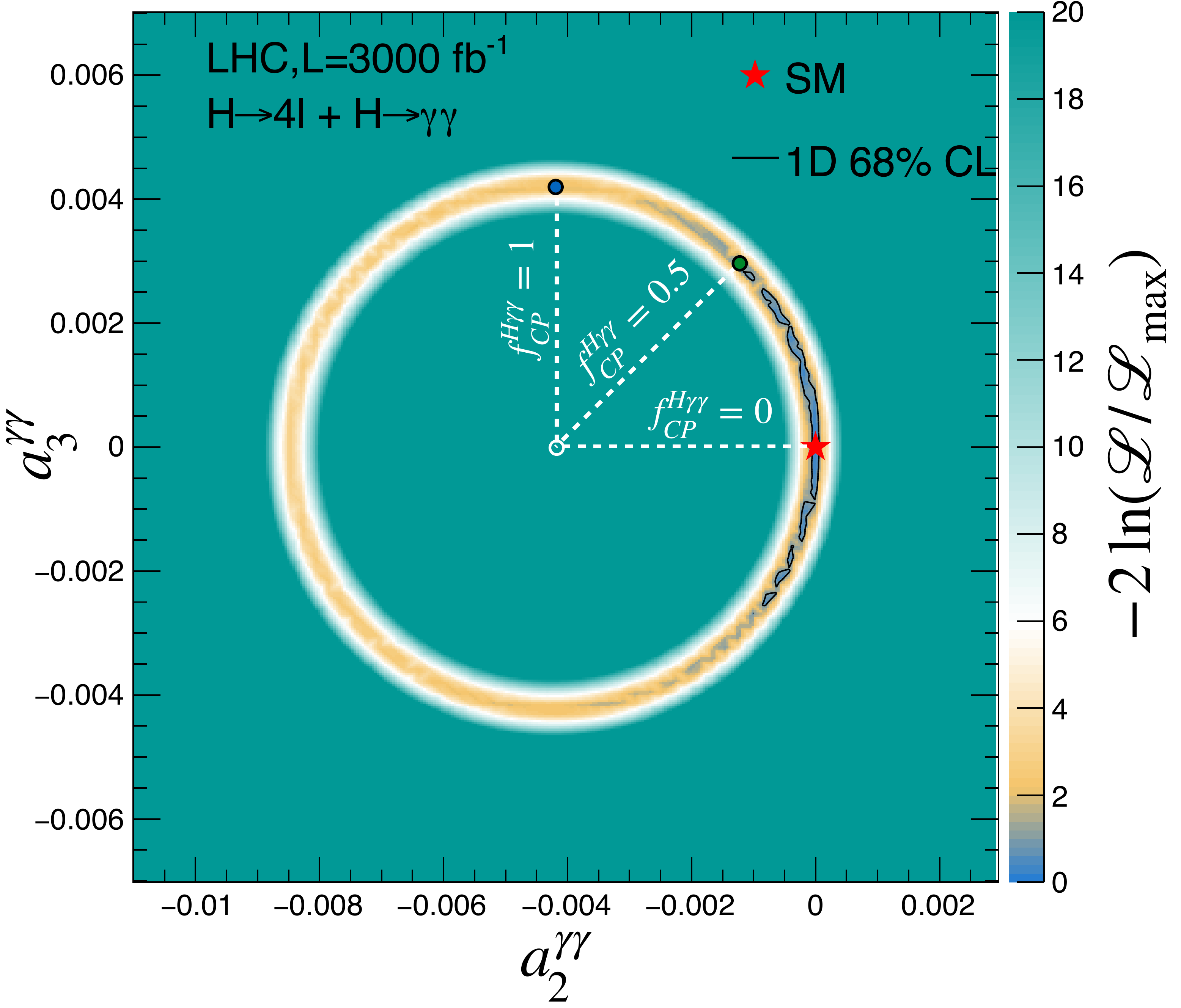}
~~~~~~~~
\includegraphics[width=0.4\textwidth]{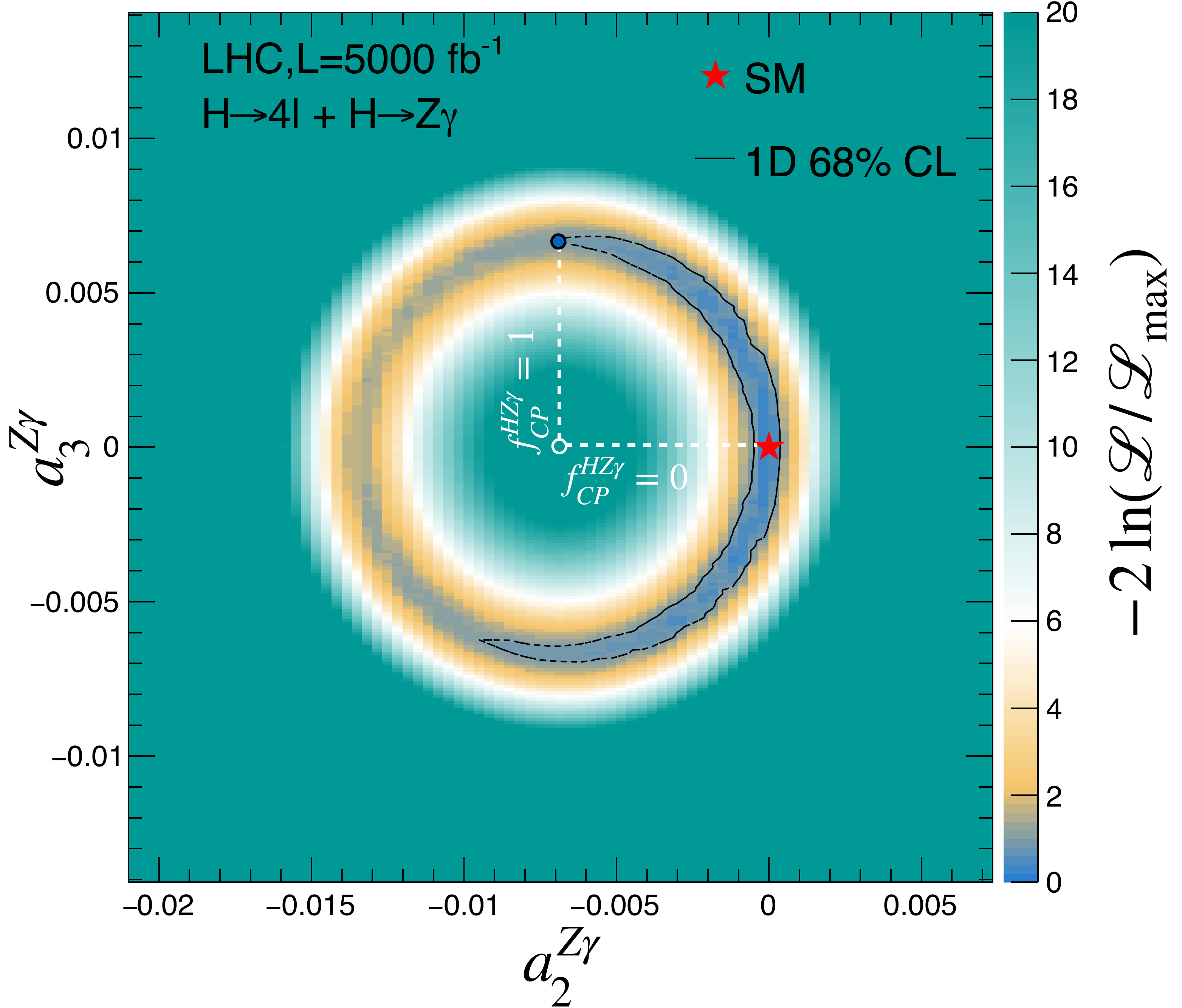}
\captionsetup{justification=centerlast}
\caption{
Expected two-dimensional constraints on
($a_2^{\gamma\gamma}$, $a_3^{\gamma\gamma})$ (left), and $(a_2^{Z\gamma}, a_3^{Z\gamma})$ (right)
using Eq.~(\ref{eq:ratio-Vgamma}) and the HL-LHC projection of analysis of the 
$H\to\gamma\gamma$, $H\to Z\gamma$, $H\to 4\ell$, VBF, and $VH$ 
channels with 3,000\,fb$^{-1}$ (left) and 5,000\,fb$^{-1}$ (right) following the study from Ref.~\cite{Davis:2021tiv}. 
}
\label{fig:pp_2D}
\end{figure}

The constraints on $(a_{2}^{\gamma\gamma}, a_{3}^{\gamma\gamma})$ and $(a_{2}^{Z\gamma}, a_{3}^{Z\gamma})$
from the HL-LHC measurements of the $H\to\gamma\gamma$ and $H\to Z\gamma$ decay rates, assuming that production
rates can be constrained in the global analysis of the \Hboson data to a good enough precision, appear as circles on
the 2D planes, as indicated in Fig.~\ref{fig:pp_2D}. These circles correspond to the fixed values of ${\rm g}^{\rm eff}_{HV\gamma}$
in Eq.~(\ref{eq:ratio-Vgamma}). The centers of the circles are at $(-a_{2}^{V\gamma, \rm SM},0)$.
All points on a circle of a given radius have equal probability, and rotation around the circle can be parameterized 
with the $f_{CP}^{HV\gamma}$ value, as indicated on the graphs in Fig.~\ref{fig:pp_2D}. With the $H\to\gamma\gamma$ 
and $H\to Z\gamma$ decay rates only, the $f_{CP}^{HV\gamma}$ values are not constrained. 

It has been demonstrated in Ref.~\cite{Davis:2021tiv} that the data from $H\to 4\ell$, 
VBF, and $VH$ can resolve the points along the circles on the  $(a_{2}^{V\gamma}, a_{3}^{V\gamma})$ plane. 
While the VBF and $VH$ channels do provide information to differentiate the $CP$-odd and $CP$-even couplings, 
the dominant precision comes from the $H\to ZZ/Z\gamma^*/\gamma^* \gamma^* \to 4\ell$ process, 
and we refer to Ref.~\cite{Davis:2021tiv} for an explanation of this effect. 
A 68\% C.L. exclusion of $f_{CP}^{H\gamma\gamma}=0.5$ can be achieved with 3,000\,fb$^{-1}$
(left plot in Fig.~\ref{fig:pp_2D}), while $f_{CP}^{HZ\gamma}=1.0$ can be excluded 
with 5,000\,fb$^{-1}$ (right plot in Fig.~\ref{fig:pp_2D}). We take these as estimates of the HL-LHC
precision on $f_{CP}^{H\gamma\gamma}$ and $f_{CP}^{HZ\gamma}$, but note that a more detailed
study and incorporation of multiple production channels may improve this further. 

\clearpage

\section{Recent updates of the studies at an electron-positron collider}
\label{app:B}

{\it Contributed by Lucas~S.~Mandacar\'{u}~Guerra and Savvas Kyriacou.}
\\

In this Section, we present a feasibility study of the $CP$-odd $H\gamma\gamma$ and $HZ\gamma$ interactions 
at an $e^+e^-$ machine and revise the study of the $CP$-odd $HZZ$ interactions documented in Snowmass-2013
writeup~\cite{Dawson:2013bba} and Ref.~\cite{Anderson:2013afp}. We start with the study of the $e^+e^-\to VH$ 
production at $\sqrt{s}=250$\,GeV and 250\,fb$^{-1}$, with $H\to b\bar{b}$ and $V\to\ell\ell$. We note that with 
the $H\gamma\gamma$ and $HZ\gamma$ couplings, both $V=Z$ and $\gamma^*$ are possible. The dominant 
contribution comes from the SM $HZZ$ couplings, and in Ref.~\cite{Anderson:2013afp} it is estimated that about 1870 
events would be reconstructed. The dominant background is modeled with the process $e^+e^-\to ZZ/Z\gamma^*\to b\bar{b}\ell\ell$.
The analysis is based on the 4D parameterization of the mass-angular distributions
$(m_{\ell\ell}, \cos\theta_1, \cos\theta_2, \Phi)$ and otherwise follows a similar technique 
to that employed in HL-LHC studies in Appendix~\ref{app:A}.

\begin{figure}[h!]
\includegraphics[width=0.24\textwidth]{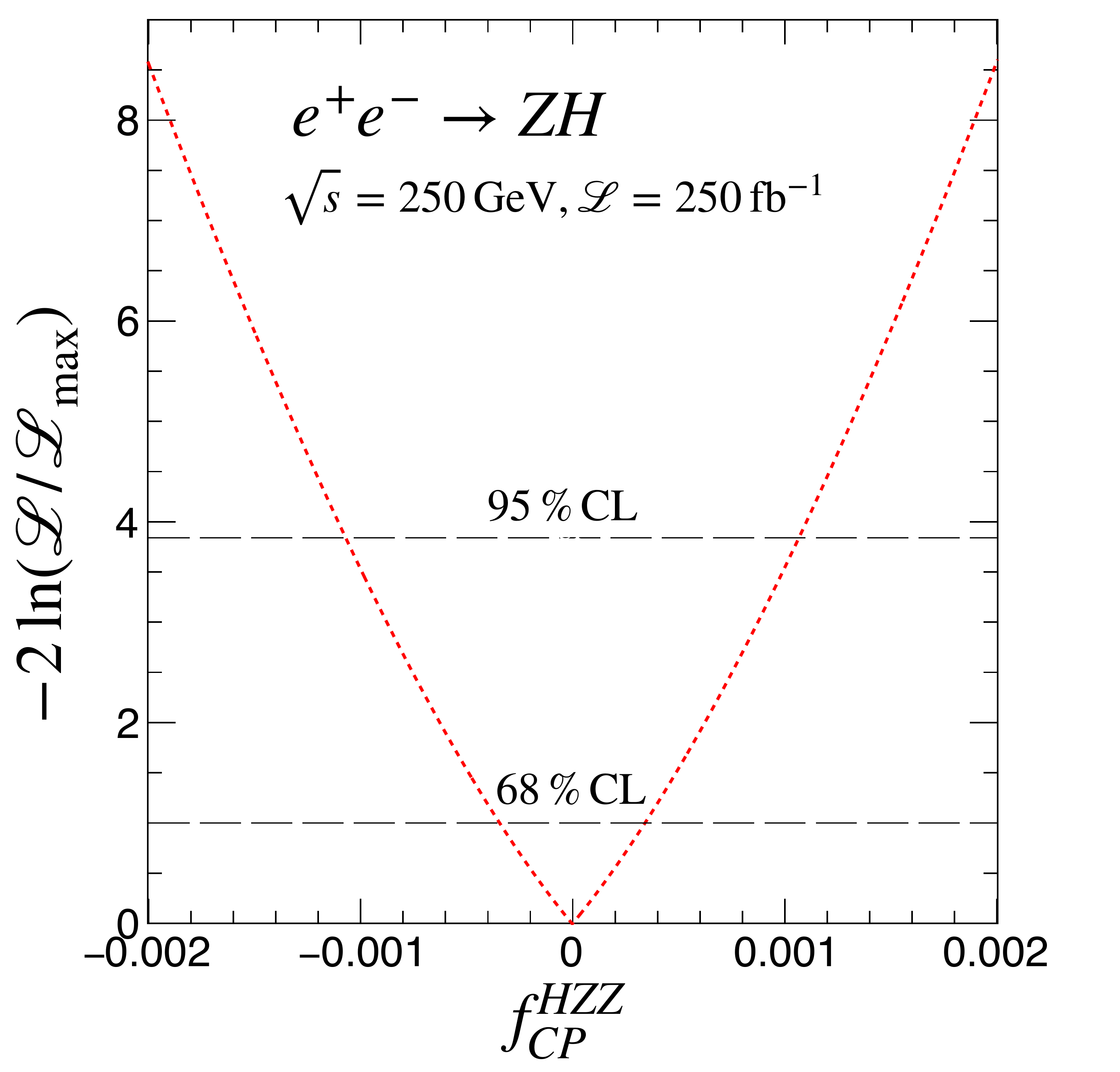}
\includegraphics[width=0.24\textwidth]{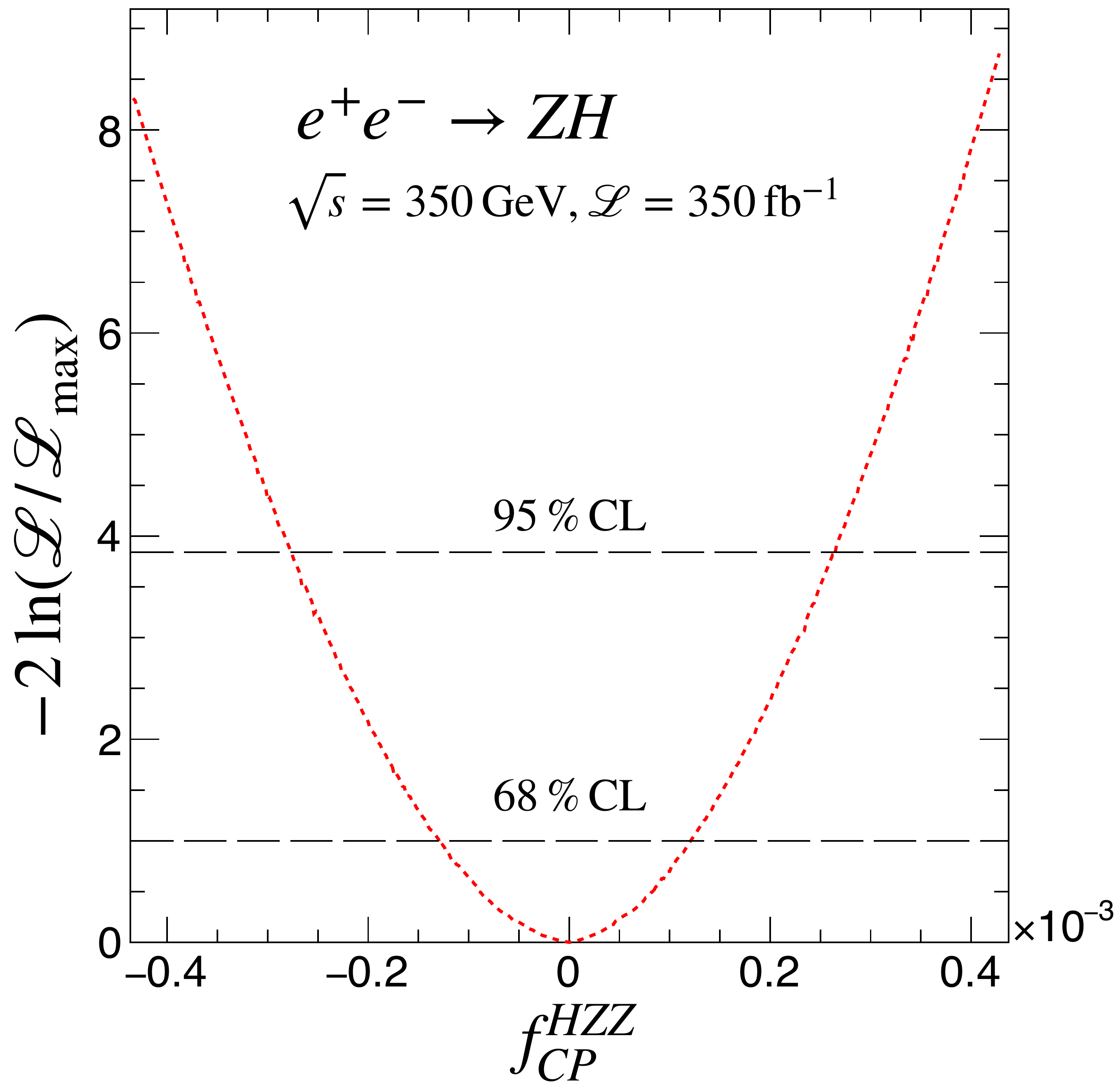}
\includegraphics[width=0.24\textwidth]{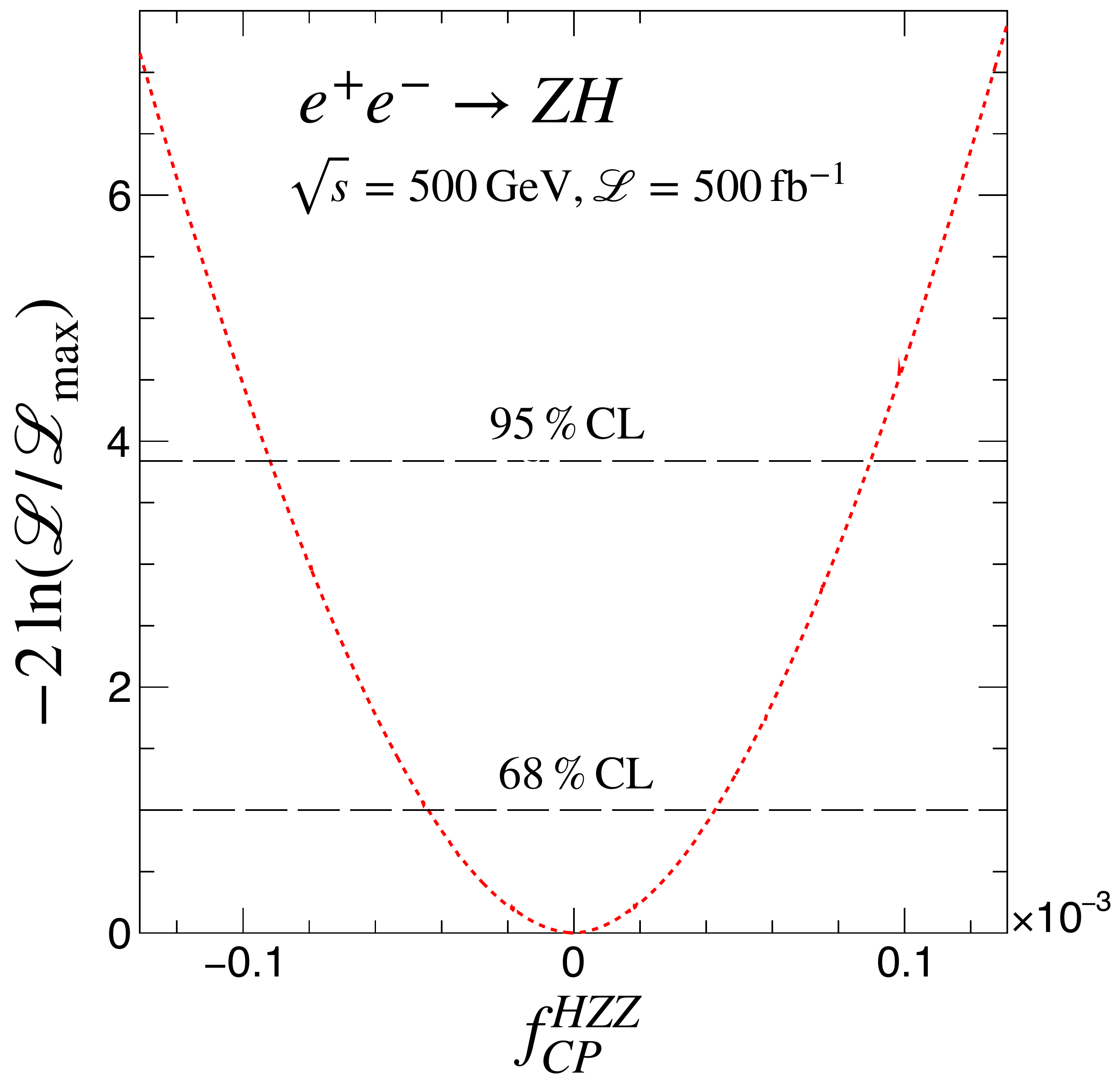}
\includegraphics[width=0.24\textwidth]{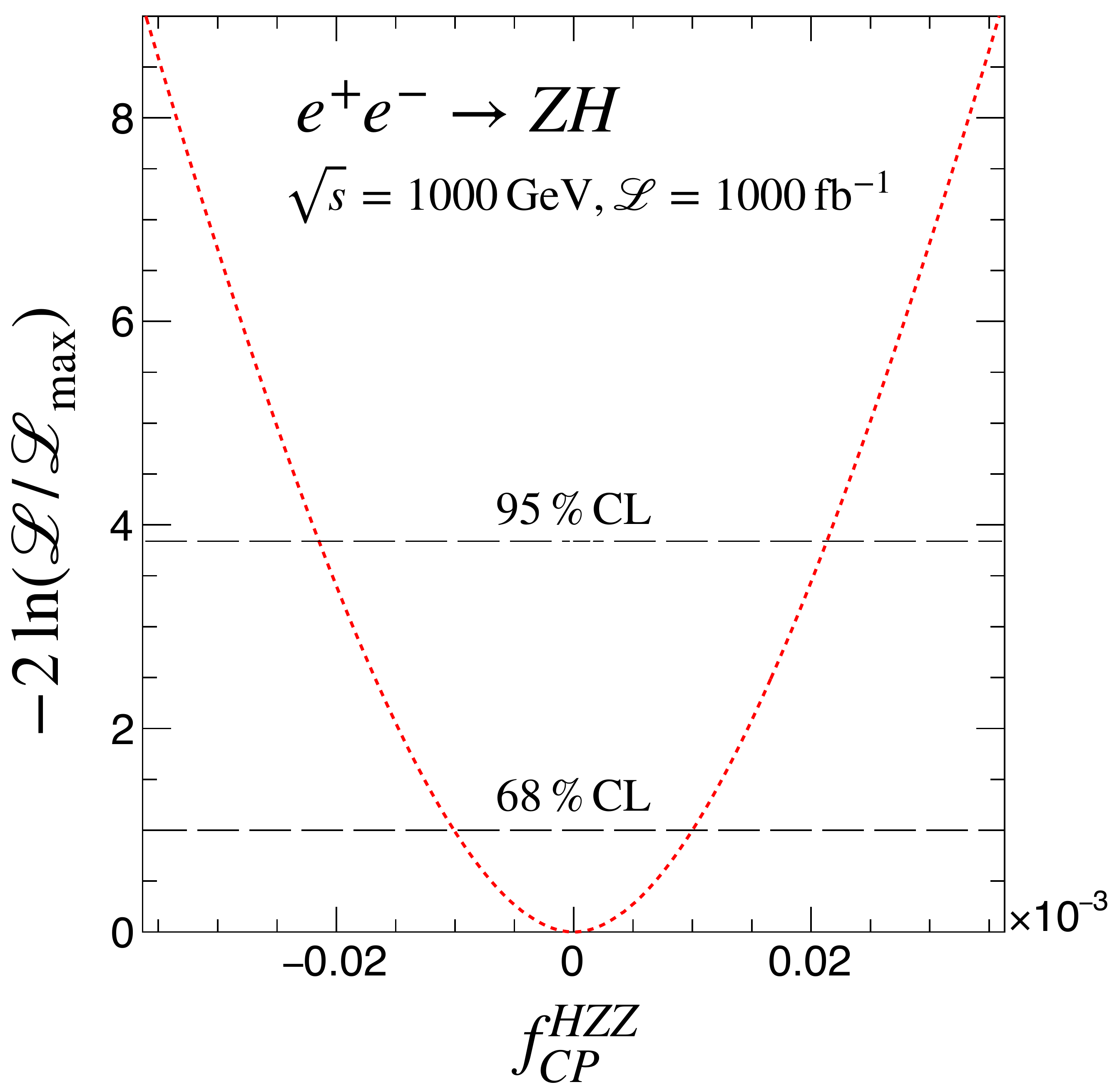}
\captionsetup{justification=centerlast}
\caption{
Expected constraints on the $f_{CP}^{HVV}$ parameter in the process $e^+e^-\to Z^*\to ZH\to \ell\ell b\bar{b}$
at four energy and luminosity scenarios.
}
\label{fig:ee_fa3}
\end{figure}

First, we reproduce the feasibility study of the $f_{CP}^{HZZ}$ parameter and find results consistent with those 
reported in Ref.~\cite{Anderson:2013afp}. The expected constraints at four energy and luminosity scenarios 
are shown in Table~\ref{table-ee-revised} and the likelihood scans are shown in Fig.~\ref{fig:ee_fa3}.
%
\begin{table}[ht]
\renewcommand{\arraystretch}{1.3}
\captionsetup{justification=centerlast}
\caption{
List of expected precision (at 68\% C.L.) of ${CP}$-sensitive measurements of the parameter $f_{CP}^{HVV}$
defined in Eq.~(\ref{eq:fractions}) in the process $e^+e^-\to Z^*\to ZH\to \ell\ell b\bar{b}$ in several energy and 
luminosity scenarios.
Also shown are expected constraints (at 68\% C.L.) on the $f^{V\gamma}$ parameters in the same process,
which are expressed as fractions of the $a_2^{V\gamma}$ and $a_3^{V\gamma}$ contributions combined 
in the $H\to 2e2\mu$ decay cross sections, 
where the most likely values of $f^{\gamma\gamma}=0.0016$ and $f^{Z\gamma}=0.0050$ were generated,
corresponding to $a_{2}^{\gamma\gamma, \rm SM}$ and $a_{2}^{Z\gamma, \rm SM}$. 
Only the $f_{CP}^{Z\gamma}$ parameter in the last energy and luminosity scenario 
allows a non-trivial constraint at 68\% C.L., as indicated in the last column. 
}
\vspace{-0.4cm}
\begin{center}
\begin{tabular}{|ll|c|c|c|c|c|}
\hline\hline
E (GeV) & ${\cal L}$ (fb$^{-1}$) & $f_{CP}^{HVV}$  & $f^{\gamma\gamma}$  & $f^{Z\gamma}$ & $f_{CP}^{\gamma\gamma}$  & $f_{CP}^{Z\gamma}$   \\
\hline\hline
250 & 250 & $\pm3.4\!\cdot\!10^{-4}$ & $<0.144$  & $<0.234$ & -- & -- \\
250 & 2,500 & $\pm3.9\!\cdot\!10^{-5}$ & $<0.037$  & $<0.079$ & -- & -- \\
350 & 350 & $\pm1.2\! \cdot\!10^{-4}$ & $<0.058$  & $<0.088$ & -- & -- \\
350 & 3,500 & $\pm2.9\! \cdot\!10^{-5}$ & $<0.016$  & $<0.032$  & -- & -- \\
500 & 500 & $\pm4.3\!\cdot \!10^{-5}$ & $<0.028$  & $<0.039$  & -- & -- \\
500 & 5,000 & $\pm1.3\!\cdot \!10^{-5}$ & $<0.009$  & $<0.016$  & -- & -- \\
1,000 & 1,000 & $\pm1.0\!\cdot\!10^{-5}$ & $<0.009$  & $<0.014$  & -- & -- \\
1,000 & 10,000 & $\pm3.0\!\cdot\!10^{-6}$ & $<0.004$  & $0.0050^{+0.0026}_{-0.0028}$  & -- & $\pm0.96$ \\
\hline
\hline
\end{tabular}
\end{center}
\label{table-ee-revised}
\end{table}
%
In addition, we present expected constraints with luminosity ten times larger than that considered
in Snowmass-2013 studies, which is consistent within a factor of two with the more recent 
collider scenarios considered in Snowmass-2022 studies, as outlined in Ref.~\cite{Dawson:2022zbb}. 
This indicates that the scaling with luminosity is close to linear at the lower
energy scenario, while it is in between the linear and square root at higher energies.
This luminosity dependence can also be seen in Fig.~\ref{fig:ee_fa3}. 
The new expectation at 250\,GeV is about a factor of 2 tighter than that obtained 
for Snowmass-2013~\cite{Dawson:2013bba, Anderson:2013afp}. However, the previous expected constrains 
at 68\% C.L. were obtained the assumption of a non-zero ${CP}$ violation near the threshold of discovery. 
The new constraints are obtained under the assumption of null ${CP}$ violation, to be consistent with 
other studies performed since then. The differences at other energies are even smaller. 
We also take our study one step further and obtained constraints in two ways: with all other 
${CP}$-even couplings (with the exception of the tree-level SM coupling, which is always left unconstrained)
either constrained to zero, as expected in the SM, or left unconstrained in the fit. The list of ${CP}$-even couplings 
can be found in Eq.~(\ref{eq:fullampl-spin0}) and includes $a^{HZZ}_{1}$, $a^{HZZ}_{2}$, and two other couplings
$\kappa^{HZZ}_{1}$ and $\kappa^{HZ\gamma}_{2}$, which correspond to the higher-order $q^2$ terms 
in expansion of $a^{HVV}_{1}$. The expected constraints in Table~\ref{table-ee-revised} do not
differ within quoted precision either with or without the ${CP}$-even anomalous couplings floated
in the fit. 

\begin{figure}[b!]
\includegraphics[width=0.34\textwidth]{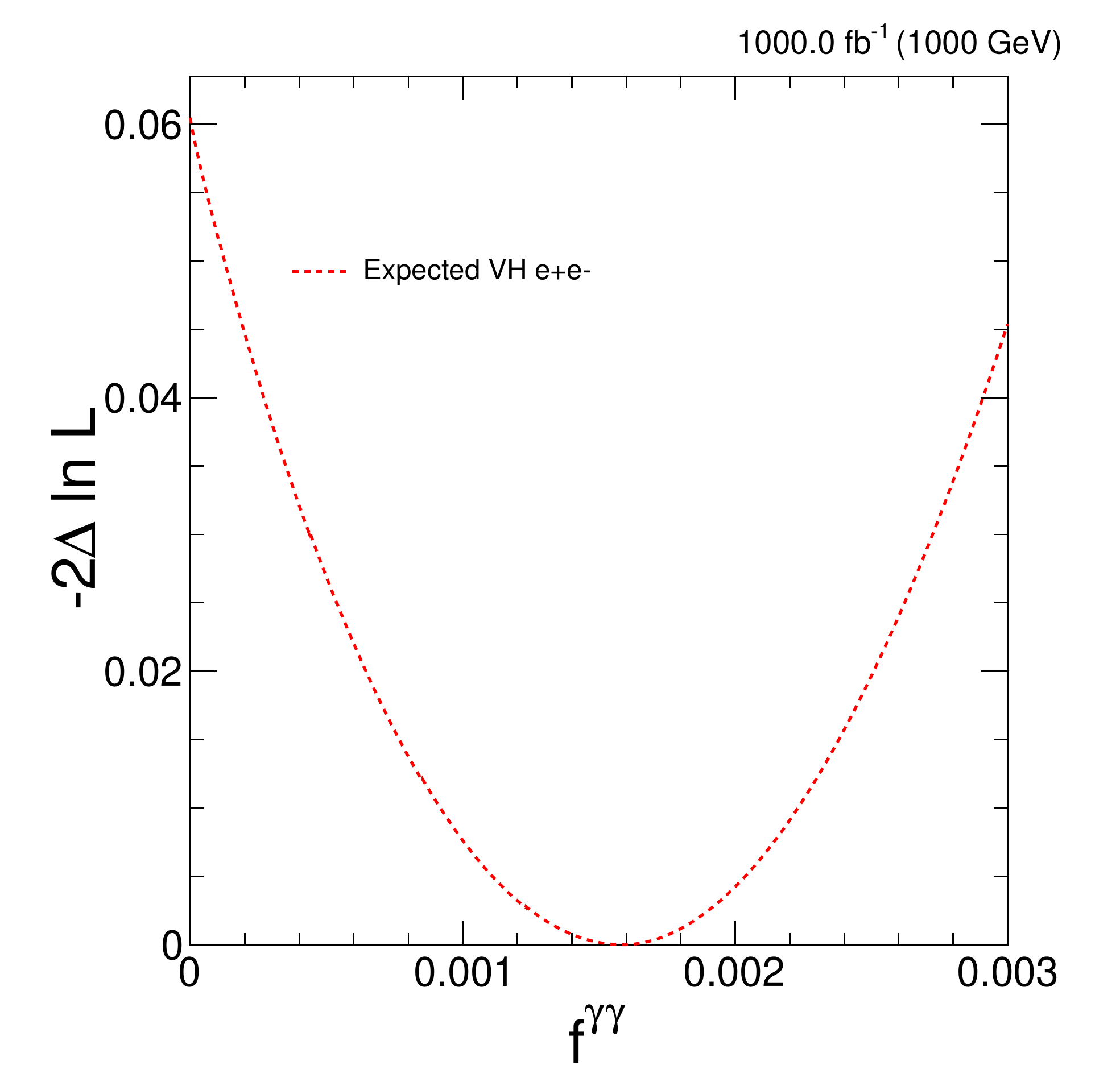}
~~~~~~~~
\includegraphics[width=0.34\textwidth]{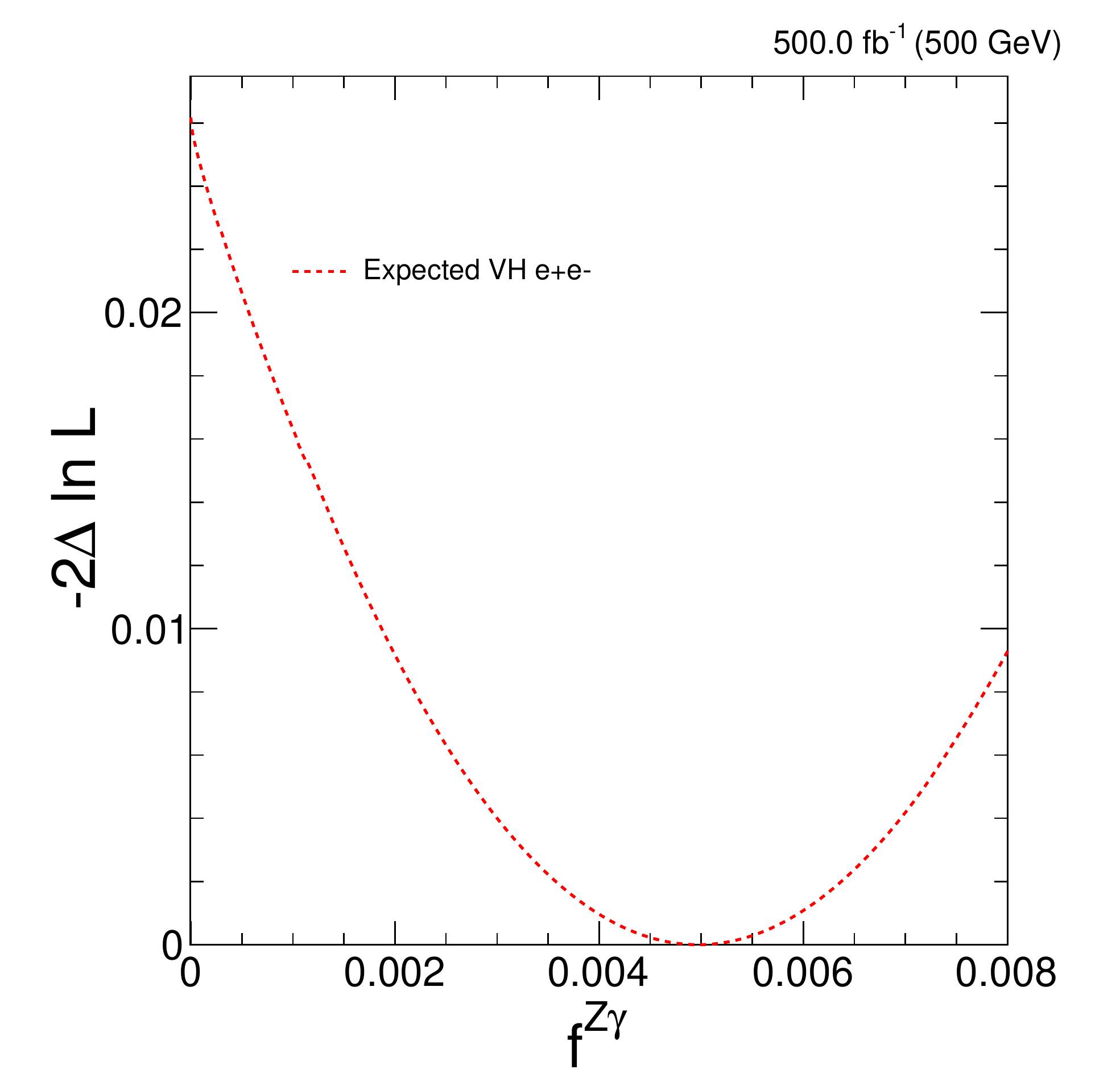}
\captionsetup{justification=centerlast}
\caption{
Examples from Table~\ref{table-ee-revised}
of expected constraints on the $f^{V\gamma}$ parameters, where the most likely values are
$f^{\gamma\gamma}=0.0016$ and $f^{Z\gamma}=0.0050$,
in the $e^+e^-\to Z^*\to ZH\to \ell\ell b\bar{b}$ process in two energy and luminosity scenarios.
}
\label{fig:ee_fa2}
\end{figure}

Then, we turn to the prospect of the 
$(a_{2}^{\gamma\gamma}, a_{3}^{\gamma\gamma})$ and $(a_{2}^{Z\gamma}, a_{3}^{Z\gamma})$
measurements in the $e^+e^-\to VH$ production. 
We can already point out that using the $a_{2}^{\gamma\gamma, \rm SM}$ and  $a_{2}^{Z\gamma, \rm SM}$
values, quoted in Appendix~\ref{app:A}, one can expect only about 0.1 and 2 events, respectively, if these
are the only contributions to the $HVV$ production amplitude in the process
$e^+e^-\to Z^*\to ZH\to \ell\ell b\bar{b}$ at $\sqrt{s}=250$\,GeV and 250\,fb$^{-1}$.
With such a small contribution, it is not feasible to expect strong constraints
on the photon couplings. Nonetheless, the full study with the 4D likelihood fit is essential to take into 
account the effects of interference of the photon couplings with the dominant SM tree-level $HZZ$
contribution. This interference is not very strong in the case of the $H\gamma\gamma$ couplings
due to very different $m_{\ell\ell}$ spectra. 
We parameterize the $a_{2}^{\gamma\gamma}, a_{3}^{\gamma\gamma}, a_{2}^{Z\gamma}, a_{3}^{Z\gamma}$
contributions in terms of four parameters: two fractions $f^{\gamma\gamma}$ and $f^{Z\gamma}$, 
expressed as the $a_2^{V\gamma}$ and $a_3^{V\gamma}$ contributions combined in the $H\to 2e2\mu$ decay 
cross sections, and two $f_{CP}^{\gamma\gamma}$ and $f_{CP}^{Z\gamma}$ parameters defined 
in Eq.~(\ref{eq:fCP}) for $H\to V\gamma$. 
Examples of the fits for $f^{\gamma\gamma}$ and $f^{Z\gamma}$ are shown in Fig.~\ref{fig:ee_fa2},
and expected upper limits in eight scenarios are shown in Table~\ref{table-ee-revised}.
We conclude that there is not enough sensitivity to isolate the 
$H\gamma\gamma$ and $HZ\gamma$ contributions with the rates generated by the 
$a_{2}^{\gamma\gamma, \rm SM}$ and $a_{2}^{Z\gamma, \rm SM}$ couplings. 
Therefore, constraints on $f_{CP}^{HZ\gamma}$ and $f_{CP}^{H\gamma\gamma}$ 
are not feasible if these contributions are comparable to the SM expectation. 
Only the $f_{CP}^{Z\gamma}$ parameter at $\mathrm{E}=1$\,TeV and  ${\cal L}=10$\,ab$^{-1}$
allows a non-trivial constraint at 68\% C.L., as indicated in the last column of Table~\ref{table-ee-revised},
but still with essentially 100\% uncertainties. 
Of course, should there be anomalously large $f^{\gamma\gamma}$ or $f^{Z\gamma}$,
one could isolate $CP$-odd contributions with a relatively high precision, but such a scenario 
is excluded by the rates of the $H\to{\gamma\gamma}$ or $H\to{Z\gamma}$ processes.
Nonetheless, we note that one will start approaching sensitivity to the SM rate faster with the higher-energy 
collider scenarios, as indicated in Table~\ref{table-ee-revised}.

Expected constraints presented in Table~\ref{table-ee-revised} assume no beam polarization in $e^+e^-$
collisions. At a linear collider, $(+80\%, -30\%)$ polarization of the $(e^-,e^+)$ beams is proposed, which 
may further improve precision of certain measurements~\cite{Dawson:2022zbb}. Therefore, we have 
repeated the feasibility study in Table~\ref{table-ee-revised} with such polarization. Generally, there are 
improvements mainly because the cross section of the background process $e^+e^-\to ZZ/Z\gamma^*\to b\bar{b}\ell\ell$
is reduced (by approximately 30\%), while cross sections of the signal $VH$ processes generated by the 
$HZZ$, $HZ\gamma$, and $H\gamma\gamma$ couplings are increased (by approximately 8\%, 22\%, and 24\%,
respectively). However, the small, positive interference between the amplitudes generated by the 
$a_{1}^{ZZ}$ and $a_{2}^{Z\gamma}$ couplings with unpolarized beams becomes large and negative with the beam polarization, 
which may compensate for such an effect in the measurement of the $f^{Z\gamma}$ parameter. 
Such large negative interference with beam polarization is stronger at lower energies.
With beam polarization, we find less than $10\%$ improvement
in expected constraints on $f_{CP}^{HZZ}$ quoted in Table~\ref{table-ee-revised}.
While we see larger variation in expected constraints on $f^{\gamma\gamma}$ and $f^{Z\gamma}$, 
due to the above effects, the picture remains qualitatively the same. 
For example, the expected constraint on $f_{CP}^{Z\gamma}$ at 68\% C.L. is improved by 12\% in the 
$\mathrm{E}=1$\,TeV and ${\cal L}=10\,\mathrm{ab}^{-1}$ scenario. 

While we have not performed a study of the $CP$-odd $H\gamma\gamma$ and $HZ\gamma$ interactions 
in VBF production $e^+e^-\to e^+e^-(ZZ/Z\gamma^*/\gamma^*\gamma^*)\to e^+e^- H$, we expect a conclusion 
similar to that in $e^+e^-\to VH$ by analogy with the HL-LHC studies in Appendix~\ref{app:A} and 
Ref.~\cite{Davis:2021tiv}. In the latter study, it was found that the $H\to ZZ/Z\gamma^*/\gamma^* \gamma^* \to 4\ell$ decay
process is more powerful in constraining the photon couplings than both the VBF and $VH$ processes, 
because the preferred range of $q^2$ in these processes leads to the $\gamma^*$ going far off-shell. 
This is the reverse of the situation with the $HZZ$ couplings, which are better constrained in production,
and where an increase in the collider energy $\sqrt{s}$ brings a benefit. 
It is possible to study the $H\to 4\ell$ process in the $e^+e^-$ production, but the expected number of events
is only about 7 at $\sqrt{s}=250$\,GeV and 250\,fb$^{-1}$, which is too small for a study. 
In a clean $e^+e^-$ environment, one could consider the hadronic decays of the $Z$ bosons in the $H\to ZZ$ process, 
but the full number of $H\to ZZ$ events of about 1600 is still much smaller than the expected number 
of $H\to 4\ell$ events at the HL-LHC. We make a preliminary conclusion that most likely it will not be 
feasible to constrain the $CP$-odd photon couplings $a_3^{V\gamma}$ of the \Hboson 
with a precision comparable to the $CP$-even contribution to the decay process $a_2^{V\gamma,\rm SM}$
at an $e^+e^-$ collider with parameters outlined in Table~\ref{table-cpscenarios}.
Nonetheless, we encourage a dedicated study of the VBF process to confirm this expectation.



\section{EDM constraints}
\label{app:C}

{\it Contributed by Wouter Dekens.}
\\

The $CP$-odd Higgs couplings not only appear in processes directly involving the Higgs boson, but also affect low-energy precision experiments through loop diagrams. Measurements of the EDMs of the neutron \cite{Abel:2020pzs}, mercury \cite{Graner:2016ses}, and the ThO molecule \cite{ACME:2018yjb} set particularly stringent constraints on $CP$-violating interactions beyond the SM. 
The loop contributions to these observables have been widely considered in the context of the SMEFT, see e.g.\ Refs.\ \cite{Brod:2013cka,Brod:2022bww,Chien:2015xha,Cirigliano:2019vfc,Dekens:2013zca,Fan:2013qn}. In these analyses, the $CP$-violating SMEFT interactions are first matched onto a low-energy theory in which the heavy SM degrees of freedom have been integrated out and subsequently evolved to the QCD scale. At this scale the quark-level theory can be matched to Chiral perturbation theory, giving rise to a description in terms of $CP$-odd interactions between hadrons, photons, and electrons, which can then be used to compute the EDMs of nucleons, atoms, and molecules.

The couplings of the Higgs to gauge bosons induce the (chromo) electric dipole moments of fermions through one-loop diagrams \cite{DeRujula:1990db,Dekens:2013zca,Fan:2013qn}, while the couplings to $t$, $\tau$, and $\mu$ contribute through two-loop Barr-Zee graphs \cite{Barr:1990vd}. 
For almost all of the couplings the most relevant contributions are those to the electron EDM, which is very stringently constrained by the ThO measurement. The exception is $f_{CP}^{H\rm gg}$, which does not induce the electron EDM and gives rise to the EDMs of the neutron and mercury instead. These different contributions have been evaluated in the SMEFT in Refs.\  \cite{Cirigliano:2019vfc} and \cite{Brod:2022bww} for the Higgs-gauge ($f_{CP}^{HVV}$) and Higgs-fermion ($f_{CP}^{Hff}$) couplings, respectively.
In this language, the $f_{CP}^{HVV}$ and $f_{CP}^{Hff}$ couplings correspond to the Wilson coefficients of dimension-six operators in the Warsaw basis \cite{Buchmuller:1985jz,Grzadkowski:2010es}, 
\begin{align}
\begin{pmatrix}
\sqrt{r_{H\rm gg}}\\\sqrt{r_{H\gamma\gamma}}\\\sqrt{r_{HZ\gamma}}\\\sqrt{r_{HZZ}}
\end{pmatrix}=
\begin{pmatrix}
\frac{12\pi}{\alpha_s}&0&0&0\\
0&-410&-120&220\\
0&130&-130&82\\
0&0.082&0.28&0.15\\
\end{pmatrix}\cdot
\begin{pmatrix}
v^2C_{H\tilde G}\\v^2 C_{H\tilde B}\\ v^2C_{H\tilde W}\\ v^2C_{H\tilde WB}
\end{pmatrix}\,,&\nonumber\\
\sqrt{r_{Htt}}= \frac{v^2 {\rm Im}\,C_{uH}^{(33)}}{y_t}\,,\quad
\sqrt{r_{Huu}}= \frac{v^2 {\rm Im}\,C_{uH}^{(11)}}{y_u}\,,\quad \sqrt{r_{Hdd}}= \frac{v^2  {\rm Im}\,C_{dH}^{(11)}}{y_d}\,,
&\nonumber\\
  \sqrt{r_{H\tau\tau}}= \frac{v^2  {\rm Im}\,C_{eH}^{(33)}}{y_\tau}\,,\quad \sqrt{r_{H\mu\mu}} = \frac{v^2  {\rm Im}\,C_{eH}^{(22)}}{y_\mu}\,,\quad \sqrt{r_{Hee}} = \frac{v^2  {\rm Im}\,C_{eH}^{(11)}}{y_e}\,,&\nonumber
\end{align}
where $r_X = \frac{f_{CP}^{X}}{1-f_{CP}^X}$, $y_f = \sqrt{2}m_f/v$, $v$ is the Higgs vacuum expectation value $v\simeq 246$ GeV, and we used the tree-level results of Ref.\ \cite{Alioli:2018ljm} to evaluate $\Gamma_{H\to VV'}^{CP \, \rm odd}$. Note that since the $f_{CP}^{HX}$ are defined through the decay rates, there is a sign ambiguity for each of the $\sqrt{f_{CH}^{HX}}$.

Using the above relations, the analyses of Refs.\ \cite{Cirigliano:2019vfc} and \cite{Brod:2022bww} can be rephrased in terms of $f_{CP}^{HVV}$ and $f_{CP}^{Hff}$, respectively. The resulting limits, assuming only one of of the couplings is nonzero at a time, are shown in Table~\ref{tab:edm}. 
The limits are dominated by the ThO measurement for all couplings apart from $f_{CP}^{H\rm gg}$, $f_{CP}^{Huu}$, and $f_{CP}^{Hdd}$, which do not induce an electron EDM and only contribute to the neutron and mercury EDMs.
Although the theoretical uncertainties related to the interpretation of the ThO measurement are small, there are significant uncertainties related to the hadronic and nuclear matrix elements that appear in the expressions for the neutron and mercury EDMs, see Refs.\ \cite{Engel:2013lsa,Chupp:2017rkp} for an overview. The table shows the constraints on the $Huu$, $Hdd$, and ${H\rm gg}$ couplings that results from varying these matrix elements within their allowed ranges, corresponding to the `Rfit' approach of Ref.\ \cite{Cirigliano:2019vfc}. In this case the dominant constraint arises from the neutron EDM. If one instead sets the matrix elements to their central values, the limits on $f_{CP}^{H\rm gg}$ and $f_{CP}^{Huu}$ ($f_{CP}^{Hdd}$) improve by a factor of $\sim 10^3$ ($ 10^2$). The most stringent limits on the Yukawa couplings are then set by the mercury EDM, while the constraints on $f_{CP}^{H\rm gg}$ from the neutron and mercury EDMs are comparable. 
The bounds in Table~\ref{tab:edm} are more stringent than those in Table~\ref{table-cpscenarios} by several orders of magnitude for the couplings of the Higgs boson to electroweak gauge bosons and the top quark. In contrast, for $f_{CP}^{H\rm gg}$ and $f_{CP}^{H\tau\tau}$, the sensitivity of the $14$ TeV LHC is comparable to the EDM constraints.

\begin{figure}
\includegraphics[width=0.3\textwidth]{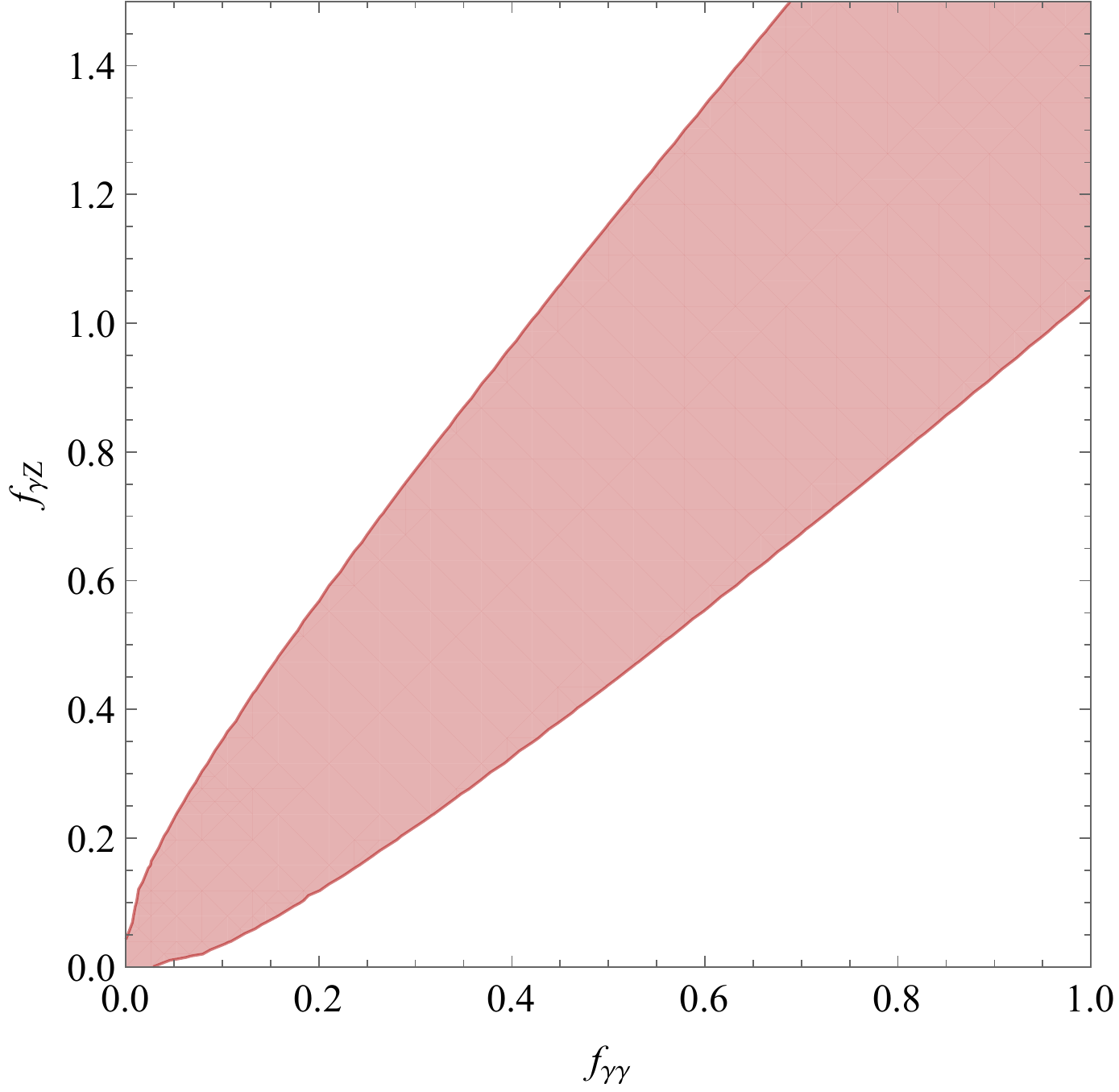}\hfill
\includegraphics[width=0.3\textwidth]{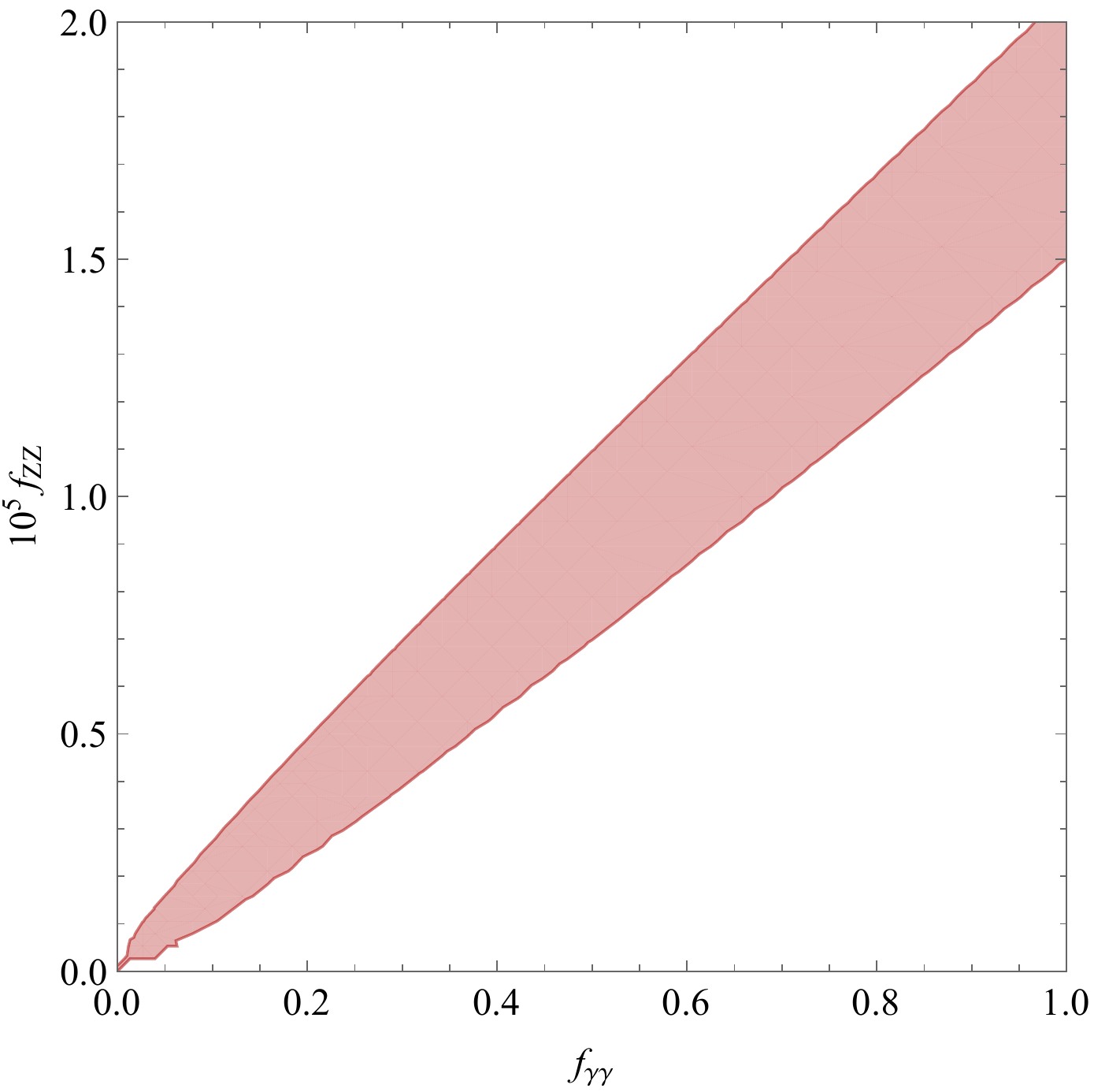} \hfill
\includegraphics[width=0.3\textwidth]{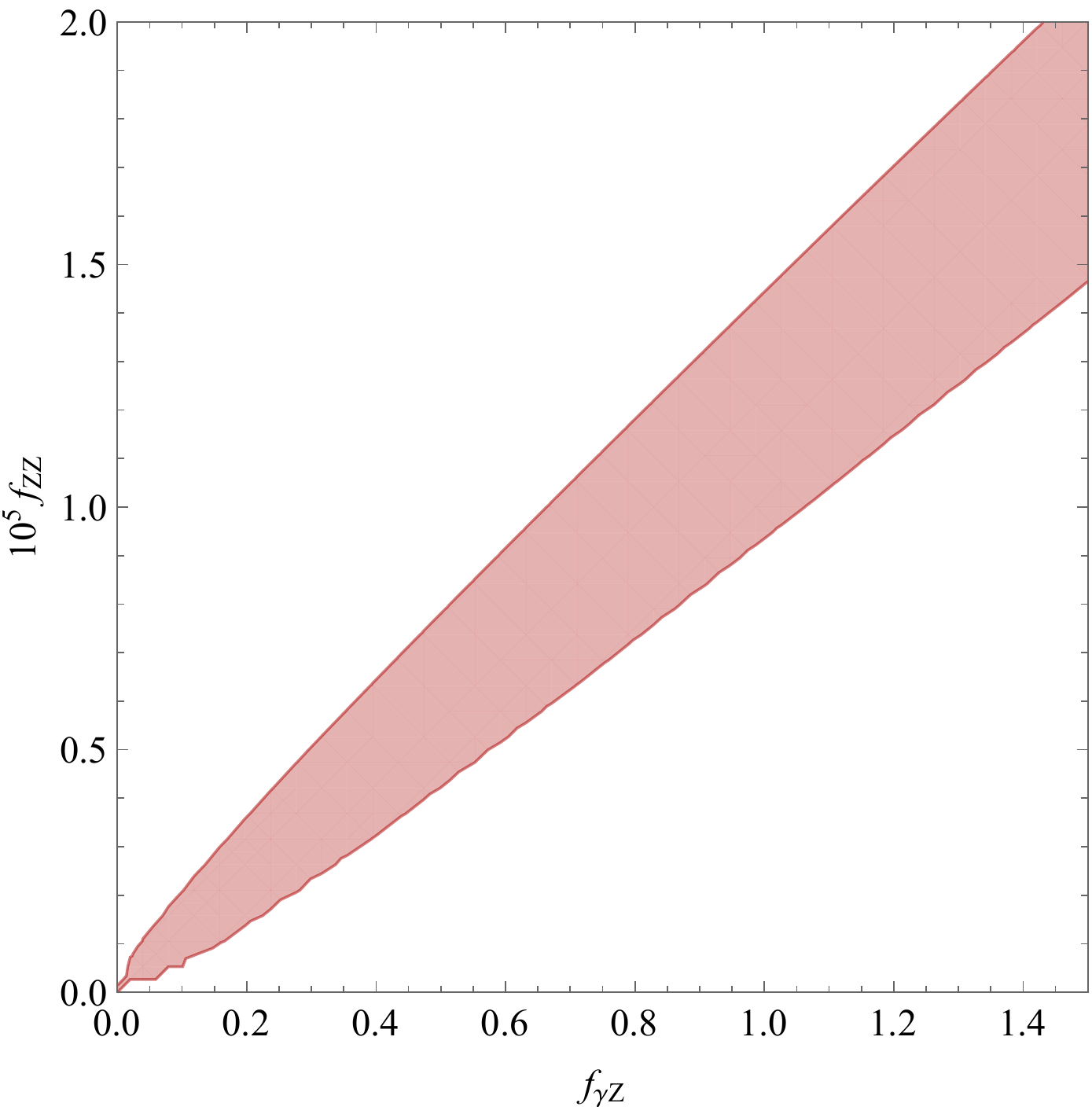}
\caption{The red shaded regions depict the parameter space allowed by EDM measurements at $90\%$ C.L.\ assuming that the $f_{CP}^{H\gamma\gamma}$,  $f_{CP}^{HZ\gamma}$, and  $f_{CP}^{HZZ}$ couplings are nonzero simultaneously.}
\label{fig:edm}
\end{figure}

Although some of the limits in Table~\ref{tab:edm} are more stringent than the projections in Table~\ref{table-cpscenarios}, they do assume that only one of the couplings is turned on at a time. However, most beyond-the-SM scenarios induce multiple operator coefficients, motivating analyses of scenarios in which several operators nonzero. As an example, we consider the case in which the three Higgs couplings to electroweak gauge bosons, $f_{CP}^{H\gamma\gamma}$ $f_{CP}^{HZ\gamma}$, and $f_{CP}^{HZZ}$, are present, with the remaining couplings set to zero. Although we in principle have measurements of the EDMs of three different systems, it turns out that they do not give enough information to constrain all three couplings, see Ref.\ \cite{Cirigliano:2019vfc} for details. As a result, there is one unconstrained linear combination of the three couplings, corresponding to a tuning of the coefficients such that the contributions to EDMs cancel. The allowed parameter space in this scenario is depicted in Fig.\ \ref{fig:edm}, where each panel shows the allowed values for two of the couplings while marginalizing over the remaining coefficient. Clearly, the couplings are allowed to be much larger than in the single-coupling analysis, in part due to the unconstrained linear combination. Nevertheless, as can be seen from Fig.\ \ref{fig:edm}, there is still a significant part of parameter space that can be excluded by EDM measurements, especially taking into account that each allowed point in these figures requires a precisely tuned value of the third coupling in order to cancel significant contributions to EDMs.

\providecommand{\href}[2]{#2}\begingroup\raggedright\endgroup

\end{document}